\documentclass[epsfig,12pt]{article}
\usepackage{graphicx}
\usepackage{amsmath}
\usepackage{makeidx}
\usepackage{stackrel}
\usepackage{amsfonts}
\usepackage{amssymb}
\usepackage{graphicx}%
\usepackage{epsfig}
\usepackage[all]{xy}
\newtheorem{theorem}{Theorem}

\makeatletter \@addtoreset{equation}{section}

\def\be{\begin{equation}}
\def\ee{\end{equation}}
\def\bea{\begin{eqnarray}}
\def\eea{\end{eqnarray}}

\newcommand{\nc}{\newcommand}
\nc{\al}{\alpha} \nc{\bib}{\bibitem} \nc{\la}{\lambda}
\nc{\C}{\mbox{\hspace{1.24mm}\rule{0.2mm}{2.5mm}\hspace{-2.7mm}
C}} \nc{\R}{\mbox{\hspace{.04mm}\rule{0.2mm}{2.8mm}\hspace{-1.5mm}
R}}

\begin{document}\title{%
\rightline{\mbox {\normalsize {WITS-CTP-144 }}\bigskip}\textbf{  Novel charges in CFT's  }}
\author{Pablo  Diaz$^{1}$\thanks{pablo.diazbenito@wits.ac.za}\\
{\small $^{1}$National Institute for Theoretical Physics, University
of Witwatersrand, South Africa}} \maketitle
\begin{abstract}
\bigskip
In this paper we construct two infinite sets of self-adjoint commuting charges for a quite general CFT. They come out naturally by considering an infinite embedding chain of Lie algebras, an underlying structure that share all theories with gauge groups $U(N)$, $SO(N)$ and $Sp(N)$.  The generality of the construction allows us to carry all gauge groups at the same time in a unified framework, and so to understand the similarities among them. The eigenstates of these charges are restricted Schur polynomials and their eigenvalues encode the value of the correlators of two restricted Schurs. The existence of these charges singles out restricted Schur polynomials among the number of bases of orthogonal gauge invariant operators that are available in the literature. \\\textbf{Keywords}: restricted Schur polynomials, finite N, Weingarten functions, conformal field theories, AdS/CFT.
\end{abstract}
\newpage
\tableofcontents
\newpage
\section{Introduction}
It has recently been shown \cite{Pablo} that finite $N$ physics for the half-BPS sector of $\mathcal{N}=4$ SYM with unitary, orthogonal and symplectic gauge groups can be studied altogether by the use of certain operators constructed exclusively from
an embedding chain of Lie algebras. \\
In this paper we generalized the operators used in \cite{Pablo} to act on generic states, not necessarily BPS, of theories with gauge group $G(N)$, which can be either $U(N),SO(N)$ or $Sp(N)$. We call these operators charges for the reasons adduced below. For theories with $G(N)$ gauge group, we find an infinite set $\{Q^{\vdash n}_{NM}| M>N\}$ of self-adjoint (under the free-field two-point function) commuting charges. The eigenstates of all these charges are restricted Schur polynomials  and analogs for $SO(N)$ and $Sp(N)$. Their eigenvalues encode (up to constants) the value of the correlator of restricted Schur polynomials, that is, the correlator of their eigenstates.\\
By exploiting the embedding chain of Lie algebras, we are able to construct another infinite set of self-adjoint commuting charges $\{Q^{\vdash \vdash n}_{NM}| M>N\}$, which also act on generic states. Their eigenstates are again restricted Schur polynomials but their eigenvalues are different from those of $\{Q^{\vdash n}_{NM}| M>N\}$. As reviewed in section \ref{prelis}, restricted Schur polynomials\cite{DSS,BCD} form a basis of gauge invariant operators and depend on three labels $(R,\mu,m)$. Charges  $\{Q^{\vdash n}_{NM}| M>N\}$ resolve the first label, whereas charges $\{Q^{\vdash \vdash n}_{NM}| M>N\}$ resolve the second. What we mean by ``resolve'' is that a number of measures with the respective charges will specify the label, and so the state. In the same spirit, labels $m$  are expected to be resolved by an analogous set of charges $\{Q^m_{NM}| M>N\}$. We will build them in a future work.\\
This paper offers the construction, properties and an interpretation of the two infinite sets $\{Q^{\vdash n}_{NM}\}$ and $\{Q^{\vdash \vdash n}_{NM}\}$. Although our first motivation is to study $\mathcal{N}=4$ SYM for its connection with string theory through the most tested AdS/CFT duality\cite{Mal,GKP,Wi}, we realize that the method applies all the same for a generic CFT in any spacetime dimension as long as the fields take values in the adjoint. And perhaps, with certain variations, this last condition might be dropped. \\

According to AdS/CFT duality, we can learn string physics from studying the dual gauge theory, we just need the appropriate dictionary. In general, for a gauge group $G(N)$ in the the gauge theory, the limit $N\to \infty$ corresponds to supergravity solutions on the string side. For finite $N$, probing physics of the gauge theory\cite{BBNS} corresponds to studying non-perturbative objects such as Giant Gravitons\cite{MST,GMT,HHI}, as well as aspects of spacetime geometry captured in the stringy exclusion principle\cite{MS}. Studying finite $N$ physics is crucial for describing such non-perturbative objects. On the one hand, it is well-established that the states of objects like D3 branes are described in the gauge theory by large operators, that is, composites of $n\sim N$ fields. On the other hand, it is known that the planar approximation breaks down at leading order for operators with $n\sim N^{1/2+\epsilon}$ with $\epsilon>0$, see \cite{GRW} for recent refinements of this threshold. This means that in order to study non-perturbative objects using the duality, one has to forget about the planar approximation and sum up all the Feynman diagrams. Finite $N$ physics develops techniques to deal with this problem.\\
Different gauge groups in $\mathcal{N}=4$ super Yang-Mills correspond to different geometries in which the string theory lives. For $U(N)$ it is known that the corresponding background is $\text{AdS}_5\times\text{S}^5$ ~\cite{Mal}, while for $SO(N)$ and $Sp(N)$ gauge groups the CFT is dual to $\text{AdS}_5\times \mathcal{R}\text{P}_5$ geometry\cite{Wi2}. Unlike string physics in $\text{AdS}_5\times\text{S}^5$, strings in the orbifold $\text{AdS}_5\times \mathcal{R}\text{P}_5$ are non-oriented. From this fact one expects the study of non-perturbative stringy physics in the orbifold to bring new insights\cite{AABF}, and they are captured by the finite $N$ physics of the CFT with gauge groups $SO(N)$ and $Sp(N)$. This was one of the motivations for the study of finite $N$ physics of orthogonal and symplectic gauge groups in \cite{CDD,CDD2}.\\
The programme of studying finite $N$ physics in the case of unitary groups was initiated in \cite{CJR} for half-BPS operators, that is, for operators built on a single complex matrix. They showed that half-BPS operators can be described by Schur operators and they demonstrated that Schur operators diagonalize the free field two-point function.
There has been a considerable progress on the study of finite $N$ physics for $U(N)$ gauge groups and, by now, we know a number of bases that diagonalize  the free field two-point function\cite{DSS,BCD,KR,BHR1,BHR2,K1,K2}. Orthogonal operators for the $Sl(2)$ sector of the theory, which involves gauge fields and their derivatives, and the action of the dilatation operator on them has been studied in \cite{DDS}. Fermion together with boson fields have been treated in \cite{DDN}, where an orthogonal restricted Schur basis for the whole $\frac{1}{8}$-BPS sector was found and the action of the dilatation operator on them described. Quiver gauge theories for unitary groups and the problem of counting gauge invariants operators in them has been recently studied\cite{DKN,PR}. We also know how to diagonalize the one-loop dilatation operator\cite{DMP,CDJ} for certain (large) operators dual to Giant Gravitons\cite{BBFH,CDL,DKS,DDGM, DR}. The diagonalization of the one-loop dilatation operator has provided new integrable sectors in the non-planar regime, with the spectrum of the dilatation operator reduced to that of decoupled harmonic oscillators which describe the excitations of strings attached to Giants\cite{DKS,DDGM,DR,DGM, DGM2}.   \\
The  programme of studying finite $N$ physics in the case of orthogonal and symplectic groups was initiated in \cite{AABF}. A detailed study of the planar spectral
problem of  $\mathcal{N}=4$ super Yang-Mills with gauge groups $SO(N)$ and $Sp(N)$ was carried
out in \cite{CKZ}. In order to tackle the non-planar regime in future works, exact correlators  of the orthogonal basis (which is the analog of the Schur basis in $U(N)$)  of half-BPS operators have been found in \cite{CDD,CDD2}, and also in the $\frac{1}{4}$-BPS sector\cite{GK1,GK2}. Recently, non-local operators for orthogonal and symplectic groups have also been considered \cite{FGT}. \\

 As said at the beginning, this paper deals with the construction of the set of charges $\{Q^{\vdash n}_{NM}\}$ and $\{Q^{\vdash \vdash n}_{NM}\}$ that appear in generic CFT's with unitary, orthogonal or symplectic gauge groups. The construction of those charges comes almost  exclusively from the infinite chain of Lie algebra embeddings 
\begin{equation}\label{embd}
\mathfrak{g}(1)\hookrightarrow \mathfrak{g}(2)\hookrightarrow \cdots
 \end{equation}
We construct the charges by taking this embedding to last consequences. Indeed, the only extra input is to force the charges to be self-adjoint with respect to the free-field two-point function of the theory. We see in sections \ref{Qn} and \ref{Qnn} that with these conditions there is not much room for choices. On the other hand, the generality of the embedding structure (\ref{embd}) makes it possible to tackle finite $N$ physics for all classical gauge groups within a unified framework.\\
We are going to summarize the main properties of these charges which, together with their construction, are the main results of this paper.
\begin{itemize}
\item The labels $N$ and $M$ refer to the way we construct the charges which also differentiate them. The charges are constructed by first performing an embedding of the fields  from $\mathfrak{g}(N)$ to $\mathfrak{g}(M)$ in (\ref{embd}) and, after some manipulation which is needed for self-adjointness, projecting the fields back to $\mathfrak{g}(N)$. See sections \ref{Qn} and \ref{Qnn}. 
\item The first surprise is that among all orthogonal bases of gauge invariant operators under the free-field two-point function, see \cite{DSS,BCD,KR,BHR1,BHR2,K1,K2}, the basis of restricted Schur polynomials is singled out by these charges. Restricted Schur polynomials are the eigenstates of the charges. Since for  CFT's there is a one-to-one correspondence between states of the theory and operators (as the operators act on the vacuum) we  will freely talk about restricted Schur polynomial states. Restricted Schur polynomials are specified by three labels: $\{R,\mu,m\}$, which have a well known group theoretical meaning in terms of representation of the group of permutations $S_n$. The point is that charges $\{Q^{\vdash n}_{NM}\}$ resolve, via their eigenvalues, the label $R$ which is a partition of $n$ if we are considering composites of $n$ fields. What we mean by ``resolve the label $R$'' is that a number of measures with charges on a certain state will determine the label $R$ of the state. Labels $\mu$, which are partitions of partitions of $n$, are resolved in the same sense by charges    $\{Q^{\vdash \vdash n}_{NM}\}$.
\item Label $m$ in the restricted Schur states is called ``multiplicity''. It also has a well-known group theoretical meaning. Labels $m$ are not resolved by charges $\{Q^{\vdash n}_{NM}\}$ or $\{Q^{\vdash \vdash n}_{NM}\}$ since their eigenvalues are the same for different labels $m$. This degeneracy is expected to be broken by a set of commuting  charges $\{Q^{m}_{NM}\}$. We leave the construction of $\{Q^{m}_{NM}\}$ for a  future work.
\item The generality of the construction of the charges, mainly of the embedding chain (\ref{embd}), makes it possible to carry all different gauge group cases at once. The construction of the charges for orthogonal and symplectic group is essentially the same as for unitary groups. The eigenstates of the charges for those gauge groups are the analogs of restricted Schur polynomials in the unitary case and their eigenvalues encode the value of the correlator of two restricted Schurs all the same.
\item The charges do not make difference upon the species that build the operators. They just concern about the values that the fields take in the Lie algebra, or more precisely, in its isomorphic matrix representation.
\item It is expected that charges $\{Q^{\vdash \vdash n}_{NM}\}$, $\{Q^{\vdash n}_{NM}\}$ and $\{Q^{m}_{NM}\}$ have an interpretation on the gravity side when the CFT we are considering is $\mathcal{N}=4$ SYM. In \cite{BCS}, regarding the half-BPS sector, it was shown that ``momenta'' which are a collection of  charges that can be expressed in terms of $\{Q^{\vdash n}_{NM}\}$, are actually encoded in asymptotic multipole moments of the dual LLM geometries. We claim in subsection \ref{inter}  that the interpretation of the charges we build in this paper are also asymptotic multipole moments of the corresponding geometries. Unfortunately, out of the BPS sector, such a description of geometries is not available yet. So, at this stage we cannot make a precise connection. 
\item Another nice surprise is that the eigenvalues of the charges encode the essential information of the correlator of two restricted Schurs. Indeed, by a simple mechanism we can recover $N$ dependence of the correlator, which is a known polynomial in $N$ of degree $n$ and depends only on the label $R$ of the state. It will be called $f_R^{G(N)}$. See the end of  subsections \ref{GM} and \ref{propertiesQnn}.
\item Regarding the eigenvalues of the charges, it is also suggesting that they admit a probabilistic interpretation. Consider the branching graph of the unitary group.  It is graded by $N$. At level $N$ we write all the irreps of $U(N)$ and we place links between irreps of consecutive levels whenever the signatures of those irreps interlace. These links form paths in the graph. Now, the eigenvalue of $\{Q^{\vdash n}_{NM}\}$  corresponding to eigenstate $R$ is the probability of starting on irrep $R$ at level $M$ on the graph and arrive at irrep $R$ at level $N$ by means of a Markov process in which one take a choice of link with probability $\frac{\text{Dim}[S,N+1]}{\text{Dim}[T,N]}$, if $S$ and $T$ are linked, in each step down.  We explain this in more detail in subsection \ref{inter}.
\end{itemize} 

The organization of the paper is as follows.\\
In section \ref{prelis} we fix the notation and give some basic background to follow the paper. The topics we talk about are structured in paragraphs. The reader is encouraged to jump to the paragraphs she considers of interest. A special mention deserves the paragraph regarding Weingarten functions, since it is a novel tool in this field. Weingarten functions are essential in the construction of the charges.\\
Sections \ref{Qn} and \ref{Qnn} contain the main results of the paper, namely the construction of the charges and their properties. Their proofs can be found in sections \ref{proofsQn} and \ref{proofQnn}.\\
One of the claims of this paper is that the charges do not make difference upon the kind of fields that build the composites. However, for simplicity, we have only used composites of bosonic fields in our proofs. In order to fill this gap we give an example with fermions in section  \ref{fermions}.\\
Section \ref{CRBCP} is an interlude between the charge constructions and their proofs. It is a technical section and can be thought as a big Lemma. The aim is to offer a characterization of restricted characters (which drive restricted Schur polynomials)  in all gauge groups by means of their algebra relations under a convolution product. In any case, section \ref{CRBCP} provides easy formulas that allow the proofs in sections \ref{proofsQn} and \ref{proofQnn} to be more general and elegant.

\section{Notation and preliminaries}\label{prelis}
This section aims to fix the notation we use all along the paper and to provide the necessary tools to follow the construction and properties of the sets of charges $\{Q^{\vdash n}_{NM}\}$ and $\{Q^{\vdash \vdash n}_{NM}\}$ for all  $M>N$ which are the main results of this paper. Such charges are going to act on generic gauge invariant operators of the theory and map them into other gauge invariant operators. The gauge invariant operators we consider are Quiral Primaries, composites of $r$ different fields $\phi_1,\dots\phi_r$  (the alphabet) from the content of the theory with abundances $\{n_1,\dots,n_r\}$. The basic bricks, the words, are multitrace monomials\footnote{We know that for orthogonal gauge groups $SO(N)$, with $N$ even, we can also built gauge invariant operators with the Pfaffian. That sector will not be dealt with in this paper.} of a total number of fields 
\begin{equation*}
n=n_1+\dots+n_r.
\end{equation*}
 For example, if we choose operators built on the scalars 2 $Z$'s and 2 $Y$'s, the possible words for unitary groups are
\begin{eqnarray*}
&&\text{Tr}(ZZYY),~ \text{Tr}(ZYZY),~ \text{Tr}(ZZY)\text{Tr}(Y),~ \text{Tr}(ZYY)\text{Tr}(Z), ~ \text{Tr}(ZZ)\text{Tr}(YY),~\text{Tr}(ZY)\text{Tr}(ZY),\nonumber \\
&&~\text{Tr}(Z)\text{Tr}(Z)\text{Tr}(YY),~ \text{Tr}(Z)\text{Tr}(Y)\text{Tr}(ZY),~\text{Tr}(Y)\text{Tr}(Y)\text{Tr}(ZZ),~ \text{Tr}(Z)\text{Tr}(Z)\text{Tr}(Y)\text{Tr}(Y).\nonumber
\end{eqnarray*}
The gauge invariant operators we are considering  are sentences with these words, that is, linear combinations of words. Charges $Q^{\vdash n}_{NM}$ and $Q^{\vdash \vdash n}_{NM}$ are maps among sentences, so they respect the field composition of the word. \\
All the fields along the paper are considered in the adjoint, as happens  in $\mathcal{N}=4$ SYM. So, they will be matrices whose algebra is isomorphic to the Lie algebra of the gauge group. With a small abuse of notation we will usually write $\phi_i\in \mathfrak{g}(N)$. Remember that the fields for orthogonal gauge groups must fulfill $\phi_i^t=-\phi_i$, whereas for symplectic groups we have $J\phi_i=-\phi_i^tJ$,  where $J$ is the standard antisymmetric matrix 
\begin{equation}\label{J}
J=\left(
\begin{array}{c|c}
 0_{N/2}&I_{N/2}  \\
\hline
-I_{N/2}&0_{N/2}  
\end{array} \right). 
\end{equation}.

Since we are going to deal with gauge invariant operators of a generic alphabet and with different gauge groups we must streamline
the notation. Let us call 
\begin{equation}\label{psi}
\Psi\equiv \phi_1^{\otimes n_1}\otimes \cdots \otimes \phi_r^{\otimes n_r}.
\end{equation}
We will need to write the components of tensor $\Psi$ explicitly. For the reasons explained below, it is convenient to write the unitary case differently from the orthogonal an symplectic. For unitary gauge groups we will write
\begin{equation}\label{psiU}
\Psi^I_J=(\phi_1)^{i_1}_{j_1}\cdots  (\phi_1)^{i_{n_1}}_{j_{n_1}} (\phi_2)^{i_{n_1+1}}_{j_{n_1+1}}\cdots  (\phi_2)^{i_{n_2}}_{j_{n_2}}\cdots\cdots   (\phi_r)^{i_{n-n_{r-1}+1}}_{j_{n-n_{r-1}+1}}\cdots  (\phi_r)^{i_{n}}_{j_{n}},
\end{equation}
whereas for orthogonal and symplectic gauge groups we set
\begin{equation}\label{psiO}
\Psi^I=(\phi_1)^{i_1i_2}\cdots  (\phi_1)^{i_{2n_1-1}i_{2n_1}} \cdots\cdots   (\phi_r)^{i_{2n-2n_{r-1}+1}i_{2n-2n_{r-1}+2}}\cdots  (\phi_r)^{i_{2n-1}i_{2n}}.
\end{equation}
Remember that for the gauge group $G(N)$, indeces $i$, or $i$ and $j$ for unitary gauge groups, run from $1,\dots,N$. 

\paragraph{Multitrace monomials.} 
The words that form our gauge invariant operators can be easily expressed in terms of $\Psi$ and permutations. We will call them
$\text{Tr}_{G(N)}(\sigma \Psi)$ and can be explicitly written in terms of (\ref{psiU}) and (\ref{psiO}) like
\begin{eqnarray}\label{multitraces}
\text{Tr}_{U(N)}(\sigma \Psi)&\equiv&\Psi^I_{\sigma(I)},\quad \sigma \in S_n,\nonumber \\
\text{Tr}_{SO(N)}(\sigma \Psi)&\equiv& \Psi^I \delta_{\sigma(I)},\quad \sigma \in S_{2n}, 
\end{eqnarray}
where $\sigma(I)=i_{\sigma(1)}i_{\sigma(2)}\dots i_{\sigma(2n)}$ and the tensor $\delta_I=\delta_{i_1i_2}\cdots\delta_{i_{2n-1}i_{2n}}$, so
\begin{equation*}
\delta_{\sigma(I)}=\delta_{i_{\sigma(1)}i_{\sigma(2)}}\cdots\delta_{i_{\sigma(2n-1)}i_{\sigma(2n)}}.
\end{equation*}
For symplectic gauge groups, where $J\phi_i J= \phi_i^t$, we will write
\begin{equation}\label{multitraceymplectic}
\text{Tr}_{Sp(N)}(\sigma \Psi)\equiv(J\Psi)^I J_{\sigma(I)},\quad \sigma \in S_{2n},
\end{equation}
where $J$ multiplies every slot of $\Psi$.\\
The relation between words and permutations is not one-to-one. Different permutations can drive the same word. In order to classify the words, and ultimately the sentences, it is important to know the symmetry under changes in the symmetric group they enjoy. We will see that these symmetries depend on the gauge group under consideration and on the precise tensor product $\Psi$.

\paragraph{Symmetries of multitrace monomials.}  
Fields $\phi_i$ can be bosonic or fermionic. If we consider bosonic fields, either scalars or the gauge field and its derivatives (derivatives of the gauge field must be considered as different scalars, see \cite{DDS}), then it is clear from (\ref{multitraces}) and (\ref{psiU}) that for unitary groups
\begin{equation}\label{simetriesUbosonic}
\text{Tr}_{U(N)}(\gamma\sigma\gamma^{-1} \Psi)=\text{Tr}_{U(N)}(\sigma \Psi),\quad \gamma\in S_{n_1}\times S_{n_2}\times \cdots \times S_{n_r}\subset S_n.
\end{equation}
The subgroup $S_{n_1}\times S_{n_2}\times \cdots \times S_{n_r}$ is called the Young subgroup, associated with $S_n$ and the partition $\lambda=(n_1,\dots ,n_r)$, and we will denote it by $S_\lambda$, so $S_\lambda\subset S_n$. Elements of $S_\lambda$ have the form $\gamma_1\circ \cdots \circ \gamma_r$. In (\ref{simetriesUbosonic})  these permutations swap the $n_i$ slots of each field $\phi_i$ in $\Psi$. An irreps $\mu$ of $S_\lambda$ is a collection of $r$ partitions $\mu=(s_1,\dots, s_r)$, where $s_i\vdash n_i$.\\
If some of the fields are fermionic, since they are Grassmann valued, their components anticommute. So, a change of slots must be accompanied with a minus sign when we change an odd number of slots. Then   (\ref{simetriesUbosonic}) turns into
\begin{equation*}
\text{Tr}_{U(N)}(\gamma\sigma\gamma^{-1} \Psi)=\text{Tr}_{U(N)}(\sigma \Psi)\text{sgn}(\gamma^f),\quad \gamma=\gamma^b\circ \gamma^f \in S_\lambda.
\end{equation*}

For orthogonal and symplectic groups the relevant symmetries are related to $S_{2n}$ instead of $S_n$ as we can see from  (\ref{multitraces}). To be precise, the symmetries will be related to subgroups of $S_{2n}$ called hyperoctahedral groups.   
Remember that any element of the hyperoctahedral group, as a subgroup of $S_{2n}$, can be seen as a $S_n$ permutation of the pairs $\{1,2;3,4;\dots;2n-1,2n\}$ plus an arbitrary flip of the elements of any pair. We will call the hyperoctahedral group $S_n[S_2]$. The symmetries of multitraces for both orthogonal and symplectic group will also be related to the Young subgroup.
In these cases, instead of $S_\lambda\subset S_n$ we will consider $S_{2\lambda}\subset S_{2n}$, where $2\lambda=(2n_1,\dots,2n_r)$, and its hyperoctahedral subgroup $S_\lambda[S_2]\subset S_{2\lambda}$.\\

For orthogonal gauge groups we have to keep in mind that the fundamental fields are matrices of $\mathfrak{so}(N)$, so they have the additional symmetry $\phi_i=-\phi_i^t$. It is easy to see that
\begin{equation*}
\Psi^{\eta(I)}=\Psi^I\text{sgn}(\eta),\quad \eta \in S_{\lambda}[S_2].
\end{equation*} 
Besides, tensor $\delta_I=\delta_{\xi(I)}$ for all $\xi\in S_n[S_2]$. So, we can see from the definition  (\ref{multitraces}) that
\begin{equation*}
\text{Tr}_{SO(N)}(\eta\sigma\xi \Psi)=\text{Tr}_{SO(N)}(\sigma \Psi)\text{sgn}(\eta), \quad \eta \in S_{\lambda}[S_2],\quad \xi\in S_n[S_2],
\end{equation*}
if all $\phi_i$ are bosonic, and if we include fermionic fields we have
\begin{equation*}
\text{Tr}_{SO(N)}(\sigma \Psi)= \text{Tr}_{SO(N)}(\eta\sigma\xi \Psi)\text{sgn}(\eta)\widetilde{\text{sgn}}(\eta_f),\quad \eta=\eta^b\circ\eta^f \in S_\lambda[S_2],~\xi \in S_n[S_2],
\end{equation*}
where the function $\widetilde{\text{sgn}}(\eta_f)$ is a 1-dimensional representation (there are four) of the hyperoctahedral group that assigns a sign to the `$S_n$' part of the permutation $\eta^f$, regardless of the flips. The role of $\eta^f$ is therefore to change the slots of fermions of the same kind, producing a minus sign if the permutation is odd. \\

For symplectic groups we perform a similar analysis keeping in mind that $J_{\eta(I)}=J_I\text{sgn}(\eta)$ for all $\eta\in S_n[S_2]$, and that $(J\Psi)^{\xi(I)}=(J\Psi^I)$ for $\xi\in S_\lambda[S_2]$. Then from (\ref{multitraceymplectic}) we see that
\begin{equation*}
\text{Tr}_{Sp(N)}(\eta\sigma\xi \Psi)=\text{Tr}_{Sp(N)}(\sigma \Psi)\text{sgn}(\xi), \quad \eta \in S_n[S_2],\quad \xi\in S_\lambda[S_2],
\end{equation*}
for bosonic fields, and as we add fermions we generalize it to
\begin{equation*}
\text{Tr}_{Sp(N)}(\sigma \Psi)= \text{Tr}_{SO(N)}(\eta\sigma\xi \Psi)\text{sgn}(\xi)\widetilde{\text{sgn}}(\xi^f),\quad \xi=\xi^b\circ \xi^f \in S_\lambda[S_2],~\eta\in S_n[S_2].
\end{equation*}

In many calculations there appear the order of the symmetric group and the subgroups we are dealing with. In order to present them clearer we try to avoid writting their explicit value. In turn we give it here:
\begin{equation*}
|S_n|=n!,\quad |S_\lambda|=n_1!n_2!\cdots n_r!, \quad |S_n[S_2]|=2^nn!,\quad |S_\lambda[S_2]|=2^n n_1!\cdots n_r!.
\end{equation*}
The fact that orthogonal and symplectic multitrace monomials have symmetry under hyperoctahedral groups (and not under symmetric groups) leads to the consideration of functions that are constant in the double coset of $S_{2n}$ with two hyperoctahedral subgroups. In fact, this kind of functions appear once and again when we deal with orthogonal and symplectic gauge groups. \\

\paragraph{Spherical functions and zonal polynomials.}
For functions that are constant on the double coset $S_n[S_2]\backslash S_{2n}/S_n[S_2]$ there is a well studied basis called `spherical functions' (see, for instance \cite{M}), they are defined as
 \begin{equation}\label{omega}
\omega_{R}(\sigma)=\frac{1}{|S_n[S_2]|}\sum_{\xi \in S_n[S_2]}\chi_{2R}(\sigma \xi), \quad \sigma \in S_{2n}\quad R\vdash n.
\end{equation}
Spherical functions enjoy the orthogonality relations inherited from characters. \\
Associated with each $\omega_{R}$ there is a symmetric polynomial $Z_{R}$ (zonal polynomial).  Zonal polynomials are the analogues of Schur functions. They are defined in terms of  spherical functions as\footnote{Remember that Schur functions are defined as $s_R( (x_1,\dots, x_N)=\frac{1}{|S_n|}\sum_{\sigma\in S_n}\chi_R(\sigma)p_\sigma (x_1,\dots, x_N)$, with  $p_\sigma$ being the power sums.}
\begin{equation}\label{zonal}
Z_{R}(x_1,\dots, x_N)=\frac{|S_n[S_2]|}{|S_{2n}|}\sum_{\rho \in S_n}2^{-l(\rho)}\omega_{R}(h_{\rho})p_{\rho}(x_1,\dots, x_N),\quad R\vdash n,
\end{equation}
where $S_{2n}\ni h_\rho=\rho (12)(34)\cdots(2n-1~2n)$, and $\rho\in S_n$ acting on the set of numbers $\{1,3,\dots,2n-1\}$. Symmetric functions $p_\rho(x_1,\dots, x_N)$ are the power sums. The function $l(\rho)$ measure the number of cycles of permutation $\rho$.     \\ 
 Zonal polynomials have  a natural `cut off', that is
$Z_{R}({\bf 1}_M)=0$ if $l(R)>M$.\\
 The specializations of zonal polynomials to 1 in their variables are 
\begin{equation}\label{zonalspecialO}
Z_{R}({\bf 1}_N)=\prod_{(i,j)\in R}(N+2j-i-1).
\end{equation} 
Alike the $U(N)$ case, there is a direct relation between specializations of zonal polynomials and combinatorial functions $f_R^{SO(N)}$, see equation (\ref{fRO}). Functions $\omega_{R}$ and the corresponding functions $\omega_\mu$ for subgroups $S_\lambda\subset S_{2n}$, as well as their associated $Z_{R}({\bf 1}_M)$ and $Z_{\mu}({\bf 1}_M)$,  will appear when dealing with orthogonal gauge groups.\\

For symplectic groups there will appear functions with the symmetry $f(\eta\sigma\xi)=f(\sigma)\text{sgn}(\eta\xi)$, where $\sigma \in S_{2n}$ and $\xi,\eta$ belonging to some hyperoctahedral subgroup. There is a well known basis for functions of this kind when $\eta,\xi\in S_n[S_2]$. They are called `twisted spherical functions' and are defined as
\begin{equation}\label{omegae}
\omega_{R}^{\varepsilon}(\sigma)=\frac{1}{|S_n[S_2]|}\sum_{\xi \in S_n[S_2]}\chi_{R\cup R}(\sigma \xi)\text{sgn}(\xi), \quad \sigma \in S_{2n},\quad R\vdash n.
\end{equation}
Note that there is a simple relation between spherical and twisted spherical functions
\begin{equation*}
\omega_R(\sigma)=\omega_{R'}^\varepsilon(\sigma)\text{sgn}(\sigma).
\end{equation*}
Associated to twisted spherical functions are also symmetric polynomials: the so-called twisted zonal polynomials, defined as
\begin{equation}\label{zonalsp}
Z'_{R}(x_1,\dots, x_N)= \frac{|S_n[S_2]|}{|S_{2n}|}\sum_{\rho \in S_n}2^{-l(\rho)}\omega^{\varepsilon}_{R}(h_{\rho})p_{\rho}(x_1,\dots, x_N), \quad R\vdash n,
\end{equation}
whose specialization is
\begin{equation}\label{twistedzonalspecial}
Z'_{R}({\bf 1}_N)=\prod_{(i,j)\in R}(2N+j-2i+1).
\end{equation}
Again, the specialization of twisted zonal polynomials gives the value of the combinatorial functions $f_R^{Sp(N)}$, as explicitly written in equation (\ref{fRsPZ}).\\

\paragraph{Restricted Schur polynomials and restricted characters}
A generic operator is a sentence, a linear combination of words. We may write it as
\begin{equation*}
\mathcal{O}(\Psi)=\sum_{\sigma}f(\sigma)\text{Tr}(\sigma \Psi).
\end{equation*}
When we don't have finite $N$ effects, namely when $n<N$, multitrace monomials form a basis of operators\footnote{When finite $N$ effects appear there is a cut-off, multitrace monomials still generate gauge invariant operators but they over-express them. Appropriate basis of operators, like  Restricted Schur polynomials, make the  cut-off natural to apply.} . It is clear that if we choose functions $f(\sigma)$ to be a basis of functions on $\mathbb{C}$ with the same symmetries as the multitraces, the operators $\mathcal{O}$ generated by $f$'s will form another basis of gauge invariant operators. The point is to look for basis which are orthogonal under the free two-point function of the theory.  The first example are  Schur polynomials driven by characters\cite{CJR}, which diagonalize the two-point function in the half-BPS sector. For broader sectors of the theory we can find different basis that make the job. Our construction of charges $Q^{\vdash n}_{NM}$ and $Q^{\vdash \vdash n}_{NM}$ singles out the restricted Schur polynomial basis.\\
In the bosonic sector of unitary gauge theories, restricted Schur polynomials where defined as\cite{DSS,BCD}
\begin{equation*}
\chi_{R,\mu,ij}(\Psi)=\frac{1}{|S_\lambda|}\sum_{\sigma\in S_n}\chi_{R,\mu,ij}(\sigma)\text{Tr}(\sigma\Psi),\quad R\vdash n,\quad \mu \text{ irrep of }S_\lambda,\quad  i,j=1,\dots, g(R,\mu),
\end{equation*}
where $\chi_{R,\mu,ij}(\sigma)$ are the so-called restricted characters. Restricted characters are a basis of functions of $S_n$ on $\mathbb{C}$ with the symmetry $\chi_{R,\mu,ij}(\gamma\sigma\gamma^{-1})=\chi_{R,\mu,ij}(\sigma)$ for $\gamma\in S_\lambda$, in accordance with the symmetry of multitraces. The construction of restricted characters is
\begin{equation*}
\chi_{R,\mu,ij}(\sigma)=\text{Tr}(P_{R\to \mu,ij}\Gamma_R(\sigma)),
\end{equation*}
where $P_{R\to \mu,ij}$ is a projector that acts on the carrier space of $R$, projects onto the irrep $\mu$ if it is subduced, and intertwines among copies of the multiplicities $i,j$. Find more details in \cite{DSS,BCD,DDGM}.\\
When we consider fermions, we also introduce inside the trace an involution \cite{DDN}. Roughly speaking,  restricted characters are found by inserting inside the trace some projectors, intertwiners and involutions acting on the carrier space of $R$. The special properties of restricted characters are then derived by results from the representation theory of symmetric groups.\\
Restricted characters for orthogonal and symplectic groups, which drive restricted Schur polynomials for those cases, have not been completely developed, see progress in this direction in \cite{GK1,GK2}. One of the claims of this paper is that charges $Q^{\vdash n}_{NM}$ and $Q^{\vdash \vdash n}_{NM}$, via their eigenstates, single out the restricted Schur polynomial basis for all the gauge groups. We could take it as a definition. However, in section  \ref{CRBCP}, by means of a convolution product, we give a characterization of restricted characters which is extensible to all the gauge groups. Although the restricted characters are not explicitly constructed, it turns out that their properties under convolution are enough to proof the main results of the paper. \\
We use a different normalization than usual for restricted characters for reasons that become clear in section \ref{CRBCP}. For example, for unitary gauge theories we will write
\begin{equation*}
b^{U}_{R,\mu,ij}(\sigma)\equiv \frac{d_R}{|S_n|}\chi_{R,\mu,ij}(\sigma).
\end{equation*}
The restricted Schur polynomials are driven by restricted characters as
\begin{eqnarray*}
\chi^{U(N)}_{R,\mu,ij}(\Psi)&=&\sum_{\sigma\in S_n}b^{U}_{R,\mu,ij}(\sigma)\text{Tr}_{U(N)}(\sigma\Psi),\nonumber \\
\chi^{SO(N)}_{R,\mu,i}(\Psi)&=&\sum_{\sigma\in S_{2n}}b^{SO}_{R,\mu,i}(\sigma)\text{Tr}_{SO(N)}(\sigma\Psi),\nonumber \\
\chi^{Sp(N)}_{R,\mu,i}(\Psi)&=&\sum_{\sigma\in S_{2n}}b^{Sp}_{R,\mu,i}(\sigma)\text{Tr}_{Sp(N)}(\sigma\Psi),
\end{eqnarray*}
where, in all cases, $R\vdash n$, $\mu$ is an irrep of $S_\lambda$ and $i,j$ are the multiplicities. Note that in the orthogonal and symplectic cases there is just one label for multiplicities. The reason for that, as well as their properties, may be seen in section \ref{CRBCP}.

\paragraph{Functions $f_R$.}
The correlator of restricted Schurs polynomials has been exactly found\cite{BCD}
\begin{equation}\label{corrrestricted}
\langle \chi^{U(N)}_{R,\mu,ij}(\Psi) \chi^{U(N)}_{S,\nu,kl}(\bar{\Psi})\rangle \propto \delta_{RS}\delta_{\mu\nu}\delta_{ik}\delta_{jl}f^{U(N)}_R,
\end{equation}
where the constant of proportion is a known function of $R$ and $\mu$. Functions $f_R^{U(N)}$ are polynomials in $N$ of the form\cite{CJR} 
\begin{equation*}
f_R^{U(N)}=\prod_{(i,j)\in R}(N+j-i).
\end{equation*}
Note that in the value of the correlator (\ref{corrrestricted}) we have dropped the spacetime dependence. The spacetime dependence of correlators in CFT's is trivial. It goes like $\langle \mathcal{O}(x) \bar{\mathcal{O}}'(y)\rangle\sim \frac{1}{|x-y|^{2\Delta}}$, where $\Delta$ is the conformal dimension of the operators. We will omit the spacetime dependence of correlators from now on. \\
It turns out that functions $f_R^{U(N)}$ can be obtained as a specialization of Schur functions to 1 in their variables
\begin{equation}\label{fRU}
 f_R^{U(N)}=\frac{1}{d_R}\sum_{\sigma \in S_n}\chi(\sigma)p_\sigma({\bf 1}_N)= \frac{|S_n|}{d_R}s_R({\bf 1}_N).
\end{equation}
It is a main result of this paper that the correlator of restricted Schurs polynomials with big label $R$ (no matter the nature of the fields they built them) are proportional to $s_R({\bf 1}_N)$, in the unitary case. This comes from application of $Q^{\vdash n}_{NM}$ as seen in section  \ref{Qn}. Also, by means of $Q^{\vdash n}_{NM}$ it is found that
\begin{eqnarray}\label{fRs}
\langle \chi^{SO(N)}_{R,\mu,i}(\Psi) \chi^{SO(N)}_{S,\nu,k}(\bar{\Psi})\rangle \propto \delta_{RS}\delta_{\mu\nu}Z_R({\bf 1}_N) \nonumber \\
\langle \chi^{Sp(N)}_{R,\mu,i}(\Psi) \chi^{Sp(N)}_{S,\nu,k}(\bar{\Psi})\rangle \propto \delta_{RS}\delta_{\mu\nu}Z'_R({\bf 1}_{N/2}).
\end{eqnarray}
Actually, the proportionality to $Z_R$ and $Z'_R$ in each case and the orthogonality in labels $R,S$ come from $Q^{\vdash n}_{NM}$, whereas  the orthogonality in labels $\mu$ and $\nu$ comes from the properties of $Q^{\vdash \vdash n}_{NM}$, see section \ref{Qnn}. Note that orthogonality in the multiplicities will come from a third operator $Q^m_{NM}$, the construction of which is left for a future work.\\
In order to unify notation we will define
\begin{equation}\label{fRO}
f_R^{SO(N)}\equiv Z_{R}({\bf 1}_N)=\prod_{(i,j)\in R}(N+2j-i-1),
\end{equation}
and 
\begin{equation}\label{fRsPZ}
f_R^{Sp(N)}\equiv Z'_{R}({\bf 1}_{N/2})=\prod_{(i,j)\in R}(N+j-2i+1),
\end{equation}
so as to write
\begin{eqnarray*}
\langle \chi^{SO(N)}_{R,\mu,i}(\Psi) \chi^{SO(N)}_{S,\nu,k}(\bar{\Psi})\rangle \propto \delta_{RS}\delta_{\mu\nu}f_R^{SO(N)} \nonumber \\
\langle \chi^{Sp(N)}_{R,\mu,i}(\Psi) \chi^{Sp(N)}_{S,\nu,k}(\bar{\Psi})\rangle \propto \delta_{RS}\delta_{\mu\nu}f_R^{Sp(N)}.
\end{eqnarray*}

\paragraph{Tensor traces.}
In the course of our computations there appear some traces of tensors built on Kronecker $\delta$ or on the matrix $J$ that we now define:
\begin{eqnarray}\label{gaugeinvariantfunctions}
\text{Tr}_{U(N)}(\sigma)&\equiv& \delta^I_{\sigma(I)}=N^{l(\sigma)},\quad \sigma \in S_n, \nonumber \\
\text{Tr}_{SO(N)}(\sigma)&\equiv& \delta^I\delta_{\sigma(I)}=N^{\text{co}(\sigma)},\quad \sigma \in S_{2n}, \nonumber \\
\text{Tr}_{Sp(N)}(\sigma)&\equiv& J^IJ_{\sigma(I)}=(-N)^{\text{co}(\sigma)}\text{sgn}(\sigma),\quad \sigma \in S_{2n}, 
\end{eqnarray}
where $l(\sigma)$ is the number of  cycles of $\sigma$, and $\text{co}(\sigma)$ is the number of rows of the coset-type of $\sigma$. The coset-type is a partition of $S_n$ that every element of $S_{2n}$ is naturally associated to. See details in \cite{M}. Indeces $i$ of $I$ run from $1$ to $N$.\\
It is clear from the definitions (\ref{gaugeinvariantfunctions}) that the traces have the symmetry
\begin{eqnarray*}
\text{Tr}_{U(N)}(\gamma\sigma\gamma^{-1})&=&\text{Tr}_{U(N)}(\sigma) \quad \gamma,\sigma\in S_n \nonumber \\
\text{Tr}_{SO(N)}(\eta\sigma\xi)&=&\text{Tr}_{SO(N)}(\sigma), \quad \sigma\in S_{2n},~~ \eta,\xi \in S_n[S_2] \nonumber \\
\text{Tr}_{Sp(N)}(\eta\sigma\xi)&=&\text{Tr}_{Sp(N)}(\sigma)\text{sgn}(\eta\xi) \quad \sigma\in S_{2n},~~ \eta,\xi \in S_n[S_2],
\end{eqnarray*}
so they can be expanded in terms of basis of functions with the same symmetry, namely characters and spherical functions. These expansions are known. They read 
\begin{eqnarray}\label{tracesexpansion}
\text{Tr}_{U(N)}(\sigma)&=&\frac{1}{|S_n|}\sum_{R\vdash n}d_R f^{U(N)}_R\chi_R(\sigma),\quad \sigma\in S_n \nonumber \\
\text{Tr}_{SO(N)}(\sigma)&=&\frac{|S_n[S_2]|}{|S_{2n}|}\sum_{R \vdash n}d_{2R}f_R^{SO(N)}\omega_{R}(\sigma),\quad \sigma\in S_{2n}  \nonumber \\
\text{Tr}_{Sp(N)}(\sigma)&=&\frac{|S_n[S_2]|}{|S_{2n}|}\sum_{R \vdash n}d_{R\cup R}f_R^{Sp(N)}\omega^\varepsilon_{R}(\sigma),\quad \sigma\in S_{2n},  
\end{eqnarray}
where we have already used the definitions $f_R^{SO(N)}\equiv Z_{R}({\bf 1}_N)$ and $f_R^{Sp(N)}\equiv Z'_{R}({\bf 1}_{N/2})$.\\
We will use these expansions, for instance, to write correlators of multitrace monomials
\begin{eqnarray}\label{correlatorsallgauge}
\langle\text{Tr}_{U(N)}(\sigma \Psi) \text{Tr}_{U(N)}(\tau \bar{\Psi})\rangle&=&\sum_{\rho\in S_\lambda}\text{Tr}_{U(N)}(\sigma^{-1}\rho\tau\rho^{-1})\text{sgn}(\rho_f)\nonumber \\
&=&\frac{1}{|S_n|}\sum_{R\vdash n}\sum_{\rho\in S_\lambda}d_R f^{U(N)}_R\chi_R(\sigma^{-1}\rho\tau\rho^{-1})\text{sgn}(\rho_f),\nonumber \\
\langle\text{Tr}_{SO(N)}(\sigma \Psi) \text{Tr}_{SO(N)}(\tau \bar{\Psi})\rangle&=&\sum_{\eta\in S_\lambda[S_2]}\text{Tr}_{SO(N)}(\sigma^{-1}\eta\tau)\text{sgn}(\eta)\widetilde{\text{sgn}}(\eta_f)\nonumber \\
&=&\frac{|S_n[S_2]|}{|S_{2n}|}\sum_{R \vdash n}\sum_{\eta\in S_\lambda[S_2]}d_{2R}f_R^{SO(N)}\omega_{R}(\sigma^{-1}\eta\tau)\text{sgn}(\eta)\widetilde{\text{sgn}}(\eta_f),\nonumber \\
\langle\text{Tr}_{Sp(N)}(\sigma \Psi) \text{Tr}_{Sp(N)}(\tau \bar{\Psi})\rangle&=&\sum_{\eta\in S_\lambda[S_2]}\text{Tr}_{Sp(N)}(\sigma^{-1}\eta\tau)\text{sgn}(\eta)\widetilde{\text{sgn}}(\eta_f)\nonumber \\
&=&\frac{|S_n[S_2]|}{|S_{2n}|}\sum_{R \vdash n}\sum_{\eta\in S_\lambda[S_2]}d_{R\cup R}f_R^{Sp(N)}\omega^\varepsilon_{R}(\sigma^{-1}\eta\tau)\text{sgn}(\eta)\widetilde{\text{sgn}}(\eta_f),\nonumber \\
\end{eqnarray}
where we have considered the general case in which  $\Psi$ also contains  fermionic fields.

\paragraph{Weingarten functions.}
In sections \ref{Qn} and \ref{Qnn} we will define the operators $\text{Av}_{MN}$ and $\text{Av}^\lambda_{MN}$ as integrals over the gauge group. The integrands are entries of the group matrices. It turns out that these integrals can be exactly computed  and lead to nice combinatorics involving symmetric functions. \\
Weingarten was the first in trying to compute them, and he succeeded for the asymptotic behaviour\cite{W}, that is, for large $N$.
Since 2003 on, these integrals have been computed for finite $N$ and for all the classical gauge groups, see \cite{C,CM,Mat}. The method for computing them has been baptized as `Weingarten calculus', and the combinatorial functions involved `Weingarten functions'. For its close relation with random matrix theory, Weingarten calculus has been widely applied in several fields of mathematics and physics, but as far as we know this is the first time it appears in our context.\\
Here we present the formulas of the Weingarten functions for the different gauge groups which will be necessary to follow the calculations of sections \ref{proofsQn} and \ref{proofQnn}.
\\
For unitary groups we will need to compute the integrals\cite{C}
\begin{equation*}
\int_{g\in U(M)}\text{d}g~g^{i_1}_{j_1}\cdots g^{i_n}_{j_n} (\bar{g})^{i'_1}_{j'_1}\cdots (\bar{g})^{i'_n}_{j'_n}=\sum_{\alpha,\beta\in S_n}(\alpha)^I_{I'}(\beta)^{J'}_J\text{Wg}^{U(M)}(\alpha\beta),
\end{equation*} 
where 
\begin{equation}\label{WgU}
\text{Wg}^{U(M)}(\sigma)=\frac{1}{|S_n|}\sum_{\substack{R\vdash n \\
l(R)\leq M}} \frac{d_R}{f_R^{U(M)}}\chi_R(\sigma) ,\quad \sigma\in S_n.
\end{equation}
The condition $l(\lambda)\leq N$ is necessary because Schur functions have a natural `cut off'. They are 0 if the of parts of $\lambda$ 
 exceeds the number of variables $N$. This would create a pole in (\ref{WgU}) and the Weingarten function would be ill-defined.\\
For orthogonal groups we will need the result\cite{CM}
\begin{equation*}
\int_{g\in O(M)}\text{d}g~g_{i_1j_1}\cdots g_{i_{2n}j_{2n}}=\frac{1}{|S_n[S_2]|^2}\sum_{\alpha,\beta\in S_{2n}}\delta_{\alpha(I)}\delta_{\beta(J)}\text{Wg}^{O(M)}(\alpha^{-1}\beta),
\end{equation*}
with
\begin{equation}\label{WgO}
\text{Wg}^{O(M)}(\sigma)=\frac{|S_n[S_2]|}{|S_{2n}|}\sum_{\substack{R\vdash n \\
l(R)\leq M}} \frac{d_{2R}}{f_R^{SO(M)}}\omega_R(\sigma) ,\quad \sigma\in S_{2n}.
\end{equation}
Note that the condition $l(\lambda)\leq N$ is also necessary because Zonal polynomials have a natural `cut off'. They are 0 if the of parts of $\lambda$  exceeds the number of variables $N$.
\\

For symplectic groups the integral on the entries reads\cite{Mat}
\begin{equation*}
\int_{g\in Sp(M)}\text{d}g~g_{i_1j_1}\cdots g_{i_{2n}j_{2n}}=\frac{1}{|S_n[S_2]|^2}\sum_{\alpha,\beta\in S_{2n}}J_{\alpha(I)}J_{\beta(J)}\text{Wg}^{Sp(M)}(\alpha^{-1}\beta),
\end{equation*}
with
\begin{equation}\label{WgSp}
\text{Wg}^{Sp(M)}(\sigma)=\frac{|S_n[S_2]|}{|S_{2n}|}\sum_{\substack{R\vdash n \\
l(R)\leq M}} \frac{d_{R\cup R}}{f_R^{Sp(M)}}\omega^{\varepsilon}_R(\sigma) ,\quad \sigma\in S_{2n},
\end{equation}
and, again,  $l(\lambda)\leq N$ needs to hold for the function to be well-defined.\\

Weingarten functions $\text{Wg}^{G(N)}(\sigma)$ are related to $\text{Tr}_{G(N)}(\sigma)$ in a nice way by means of the convolution product\footnote{The properties of restricted characters under this convolution product are studied in section \ref{CRBCP}.}
\begin{equation*}
f\star g~(\sigma)=\sum_{\alpha\in S_n} f(\alpha^{-1})g(\alpha\sigma)\quad \sigma\in S_n.
\end{equation*}
\\
It turns out that for $n<N$, Weingarten functions are the inverse of traces under this product
\begin{equation*}
\text{Wg}^{G(N)}\star \text{Tr}_{G(N)}~(\sigma)=\delta(\sigma),
\end{equation*}
whereas if we take into account finite $N$ effects
\begin{equation*}
\text{Tr}_{G(N)}\star\text{Wg}^{G(N)}\star \text{Tr}_{G(N)}~(\sigma)=\text{Tr}_{G(N)}(\sigma).
\end{equation*}
In a way, this relation is behind the fact that our operators $Q^{\vdash n}_{NM}$ and  $Q^{\vdash\vdash n}_{NM}$ have such special properties.

\section{Detailed construction of $Q^{\vdash n}_{NM}$}\label{Qn}
The construction of $Q^{\vdash n}_{NM}=\text{Proj}_{NM}\circ \text{Av}_{MN}$ is a straightforward generalization of the operators we used in \cite{Pablo}, to make them act on a generic gauge invariant operator which will a linear combination of multitraces
\begin{equation*}
\text{Tr}_{G(N)}(\sigma\Psi),
\end{equation*}
with 
\begin{equation*}
\Psi=\phi_1^{\otimes n_1}\otimes\cdots \otimes  \phi_r^{\otimes n_r}.
\end{equation*}
 For this reason we use the same notation and we review its construction with minor changes. 
\subsection{General methodology}\label{GM}
First, let us consider the infinite embedding chain
\begin{equation}\label{embebding}
\mathfrak{g}(1)\hookrightarrow \mathfrak{g}(2)\hookrightarrow \cdots
 \end{equation}
where $\mathfrak{g}=\mathfrak{u}, \mathfrak{so}~ \text{or}~\mathfrak{sp}$. So, and element $\phi_i\in \mathfrak{g}(N)$ can always be upgraded to $\phi_i\in \mathfrak{g}(M),~M>N$ by placing $\phi_i$ in the upper left of the matrix and filling the rest, up to dimension $M$, with 0's. Related to this embedding, there is a natural set of operators:
\begin{equation}\label{proj}
\text{Proj}_{NM}: \mathfrak{g}(M) \to  \mathfrak{g}(N).
\end{equation}
Operators $\text{Proj}_{NM}$  reduce the dimension of the matrices from $M$ to $N$ by killing the `extra' zeros of the
embedding. We can complete the definition of $\text{Proj}_{NM}$ by sending to 0 all elements $\phi_i\in\mathfrak{g}(M)$ such that the number of eigenvalues of $\phi_i$ which are different from 0 is greater than $N$.\\
We may easily extend the definition of $\text{Proj}_{NM}$ to act on gauge invariant operators. We will make  $\text{Proj}_{NM}$ to act as above on every $\phi_i$ of the composite operators. This way, $\text{Proj}_{NM}$ is a map that takes gauge invariant operators built on $\phi_i\in \mathfrak{g}(M)$ to gauge invariant operators built on $\phi_i\in \mathfrak{g}(N)$.\\
At this point it is useful to think of gauge invariant operators built on $\phi_i\in \mathfrak{g}(N)$ as vectors belonging to a vector space $V_N$. Then $\text{Proj}_{NM}:V_M\to V_N$,  and the free field two-point function is an inner product defined in each $V_N$.  
 It is natural to wonder about the adjoint operators of $\text{Proj}_{NM}$ with respect to this inner product. We will call them $\text{Av}_{MN}$, and they map  gauge invariant operators  built on $\phi_i\in \mathfrak{g}(N)$ into gauge invariant operators built on $\phi_i\in \mathfrak{g}(M)$. So, the two point function must fulfill
\begin{equation}\label{compatibility}
\langle\text{Av}_{MN}\mathcal{O}_N,\mathcal{O'}_M\rangle=\langle\mathcal{O}_N, \text{Proj}_{NM}\mathcal{O'}_M\rangle,
\end{equation}
for $\mathcal{O}$ and $\mathcal{O'}$ arbitrary gauge invariant operators.\\
The averaging operator can be constructed as
\begin{equation}\label{Av}
\text{Av}_{MN}\mathcal{O}_N(\Psi)=\int_{g\in G(M)} \text{d}g~ \text{Ad}_g (\mathcal{O}_N(\Psi)),
\end{equation}
where $G=U, SO ~\text{or}~Sp$, $\text{d}g$ is the Haar measure of the corresponding group and $\text{Ad}_g$ is the adjoint action of $g$ onto the algebra. The adjoint action of $\phi_i\in \mathfrak{g}(N)\hookrightarrow\mathfrak{g}(M)$ is defined  as usual:
\begin{equation}\label{usualadjointaction}
\text{Ad}_g(\phi_i)=g\phi_ig^{-1},\quad g\in G(M),
\end{equation}
where $\phi_i$ is embedded in $\mathfrak{g}(M)$. The adjoint action of $\Psi$ is defined as (\ref{usualadjointaction}) on {\it each} field $\phi_i$ of $\Psi$. If we define
\begin{equation*}
\big[g\big]=g^{\otimes n},\quad g\in G(M),
\end{equation*}
then we have
\begin{eqnarray*}
\text{Ad}_g(\Psi^I_{I'})&=&\big[g\big]^I_J \Psi^J_{J'}\big[g^\dagger\big]^{J'}_{I'},\quad g\in U(M), \nonumber \\
\text{Ad}_g(\Psi^I)&=&\big[g\big]^{IJ} \Psi_J,\quad g\in SO(M).
\end{eqnarray*}
When applied on multitrace monomials we obtain
\begin{eqnarray}\label{adjointonmulti}
\text{Ad}_g\big(\text{Tr}_{U(N)}(\sigma\Psi)\big)&=&\big[g\big]^I_J \Psi^J_{J'}\big[g^\dagger\big]^{J'}_{I'}(\sigma)^{I'}_I,\quad g\in U(M), \nonumber \\
\text{Ad}_g\big(\text{Tr}_{SO(N)}(\sigma\Psi)\big)&=&\big[g\big]^{IJ} \Psi_J\delta_{\sigma(I)},\quad g\in SO(M).
\end{eqnarray}
There is a subtle but crucial point in (\ref{adjointonmulti}). Indeces $i=1,\dots,N$ whereas $j=1,\dots,M$. In other words, we keep the original range of the traces, otherwise the adjoint action would be trivial. 
In a simple example, with just one scalar field, it would be
\begin{equation}\label{exad}
\text{Ad}_g\text{Tr}_{U(N)}(Z)=\text{Tr}_{U(N)}(gZg^{-1})\neq \text{Tr}_{U(N)}(Z),
\end{equation}
unless $M=N$, in which case, the adjoint action on gauge invariant operators is trivial.\\
As example (\ref{exad}) shows, given a gauge invariant operator $\mathcal{O}(\Psi)$, $\text{Ad}_g \mathcal{O}(\Psi)$ is in general not gauge invariant. However, 
the integral over the group restores gauge invariance. So, $\text{Av}_{MN}$, as defined in (\ref{Av}), is actually a map between gauge invariant operators.\\
Relation (\ref{compatibility}) with the definition of $\text{Av}_{MN}$ as in (\ref{Av}) was proved for the half-BPS sector and for each gauge group in \cite{Pablo}. We will give a proof for generic operators in section \ref{proofsQn}. Actually, by linearity, it will be enough to prove (\ref{compatibility}) for arbitrary multitrace monomials.\\ 
Note that correlators in (\ref{compatibility}) are in different spaces: in the LHS operators are built on elements $\phi_i\in \mathfrak{g}(M)$ whereas in the RHS $\phi_i\in \mathfrak{g}(N)$. Equation (\ref{compatibility}) shows that $\text{AV}_{MN}=\text{Proj*}_{NM},~ \forall M,N$ with respect to the free field two-point function of the theory, as we claim. It also shows the compatibility between Weingarten and Wick calculus. To see this let us consider two operators built on  $\Psi$, for example two multitrace monomials $\text{Tr}_N(\sigma \Psi)$ and  $\text{Tr}_N(\tau \Psi)$. One can compute the correlator of these to operators as usual, summing all possible Wick contractions. This way we get a result on the RHS of equation (\ref{compatibility}). Alternatively, we can upgrade $\Psi$ by upgrading every field, so that $\phi_i\in \mathfrak{g}(N)\hookrightarrow\mathfrak{g}(M)$ and go to the LHS of (\ref{compatibility}). One of the multitrace monomials keeps its structure, except for $\phi_i\in \mathfrak{g}(M)$. However, the other multitrace monomial is affected by  $\text{AV}_{MN}$ and turns into a complicated sum of multitrace monomials built on  $\Psi$, as can be seen in (\ref{Qnontraces}). The spectrum of this sum comes from the integrals involved in the definition of $\text{AV}_{MN}$, that is, from Weingarten calculus. For different $M$ we get a different sum. But as relation (\ref{compatibility}) states, all these sums must be arranged in a way so that they keep the same value for the two-point function. In this sense we say that Weingarten and Wick calculus are compatible.

Now, we will consider the composition
\begin{equation}\label{composition}
Q^{\vdash n}_{NM}\equiv\text{Proj}_{NM} \circ\text{Av}_{MN}.
\end{equation}
By construction,  $Q^{\vdash n}_{NM}$ are self-adjoint with respect to our correlators and they map gauge invariant operators built on $\phi_i\in \mathfrak{g}(N)$ into gauge invariant operators built on $\phi_i\in \mathfrak{g}(N).$\\
It is logical to wonder about the eigenvectors and the eigenvalues of (\ref{composition}). It turns out that restricted Schur polynomials are eigenvectors of $Q^{\vdash n}_{NM}$. Specifically, we will prove in section \ref{proofsQn} that\footnote{Within the context of symmetric functions, equation (\ref{eigenvectors}) first appeared in \cite{OO1,OO2} under the name of `coherence property' . Their purpose was to give a characterization of Schur functions. Although their definitions for $\text{Proj}_{NM}$ and $\text{Av}_{MN}$ are different from ours we have decided to keep their notation.}
\begin{equation}\label{eigenvectors}
Q^{\vdash n}_{NM}\chi^{G(N)}_{R,\mu,m}=\frac{f_R^{G(N)}}{f_R^{G(M)}}\chi^{G(N)}_{R,\mu,m}, \quad \forall M>N,
\end{equation}
for $G=U, SO~\text{and}~Sp$.\\
Since the eigenvalues in (\ref{eigenvectors}) are all different for different $R$'s in $\chi^{G(N)}_{R,\mu,m}$ and because $Q^{\vdash n}_{NM}$ are self-adjoint for all $M>N$, we conclude that restricted Schur operators are orthogonal in the capital label $R\vdash n$ for classical gauge groups of any rank.\\

Note that from (\ref{compatibility}) and (\ref{eigenvectors}) one can recover the precise form of the correlator of restricted Schurs up to a constant. That is, we can obtain
 \begin{equation}\label{cRortho}
\langle\chi^{G(N)}_{R,\mu,m},\chi^{G(N)}_{S,\nu,m'}\rangle=c(R,\mu,\nu,m,m')f_R^{G(N)}\delta_{RS}.
\end{equation}
The method is quite simple. From (\ref{eigenvectors}) we know that 
\begin{equation}\label{averaging}
\text{Av}_{MN}\chi^{G(N)}_{R,\mu,m}=\frac{f_R^{G(N)}}{f_R^{G(M)}}\chi^{G(N)}_{R,\mu,m}, \quad \forall M>N.
\end{equation}
Applying (\ref{compatibility})
to Schur operators and using (\ref{averaging}) we get
\begin{equation*}
\frac{1}{f_R^{G(M)}}\langle\chi^{G(M)}_{R,\mu,m},\chi^{G(M)}_{R,\nu,m'}\rangle=\frac{1}{f_R^{G(N)}}\langle\chi^{G(N)}_{R,\mu,m},\chi^{G(N)}_{R,\nu,m'}\rangle \quad \forall M>N.
\end{equation*}
This means that 
\begin{equation}\label{cR}
c(R,\mu,\nu,m,m')\equiv \frac{1}{f_R^{G(M)}}\langle\chi^{G(M)}_{R,\mu,m},\chi^{G(M)}_{R,\nu,m'}\rangle
\end{equation}
is finite and does not depend on $M$. Therefore, $c(R,\mu,\nu,m,m')$ is a number and not a polynomial in the rank of the gauge group.
Now, the orthogonality relation (\ref{cRortho})  follows from the orthogonality of restricted Schurs in the capital label $R$ and (\ref{cR}).
We may conclude that from (\ref{compatibility}) and (\ref{eigenvectors}) we recover the two point function up
to a constant. But this was expected. It is clear that both (\ref{compatibility}) and (\ref{eigenvectors}) still hold for
$k(R,\mu,m)\chi^{G(N)}_{R,\mu,m}$, so the freedom of multiplying every restricted Schur by a constant  should be reflected in the  two
point function. Relation (\ref{cRortho}) precisely reflects this arbitrariness.\\
In order to find $c(R,\mu,\nu,m,m')$ for $k(R,\mu,m)=1$  one must invariably get some result from Wick contractions.  Equation (\ref{cR}) shows that $\langle\chi^{G(M)}_{R,\mu,m},\chi^{G(M)}_{R,\nu,m'}\rangle$ is proportional to $f_R^{G(M)}$. Now,  by definition, we know that polynomials $f_R^{G(M)}$ all have coefficient 1 in the highest power of $M$. So, the value of $c(R,\mu,\nu,m,m')$  for $k(R,\mu,m)=1$ is the coefficient of the highest power of $M$ in the polynomial $\langle\chi^{G(M)}_{R,\mu,m},\chi^{G(M)}_{R,\nu,m'}\rangle$.\\

Let us summarize the logic of this construction. The starting point is to extract some information of gauge invariant operators
built on $n$ fields of $r$ different type: $\phi_1,\dots,\phi_r\in \mathfrak{g}(N)$, distributed as $n=n_1+\cdots +n_r$. We decide to fix $n$ but move on $N$. First, we realize that the algebras can be embedded as in (\ref{embebding}). Then we think of the most basic non-trivial set of operators that adapts to this embedding and
find $\text{Proj}_{NM}$. These operators map gauge invariant operators built on $\phi_i\in \mathfrak{g}(M)$ into gauge invariant operators built on $\phi_i\in \mathfrak{g}(N)$. Considering gauge invariant operators as vectors and the two-point function as the inner product of the theory, we wonder which operators are the adjoints of $\text{Proj}_{NM}$, we call them $\text{Av}_{NM}$ and find that they can be constructed as in (\ref{Av}). The fact that $\text{Av}_{MN}=\text{Proj*}_{NM}$ is shown in relation (\ref{compatibility}). Moreover, we construct a set of self-adjoint operators by composition of them in (\ref{composition}) and wonder about their eigenvectors and eigenvalues in (\ref{eigenvectors}). It turns out that restricted Schur polynomials are eigenvectors  whose eigenvalues are different for different capital labels\footnote{This explains why we chose to use the label $\vdash n$ in $Q^{\vdash n}_{NM}$.} $R\vdash n$.  From there, we conclude that restricted Schurs are necessarily orthogonal in those labels. Besides, the polynomial in $N$ behaviour of the correlators of restricted Schurs is completely fixed by the embedding, as shown in (\ref{cRortho}).\\

\subsection{Interpretation of $Q^{\vdash n}_{NM}$}\label{inter}
The objects $Q^{\vdash n}_{NM}$ are self-adjoint with respect to the free-field two point function and commute with each other for all $M>N$. When applied to half-BPS operators they completely specify the state. If we choose Schur polynomials as a basis of half-BPS operators then a number of measures of distinct  $Q^{\vdash n}_{NM}$, that is, a number of different $M$'s, will completely specify $R\vdash n$ which labels each state. When $Q^{\vdash n}_{NM}$ are applied to generic operators the appropriate bases to think of are restricted Schur polynomials $\chi_{R,\mu,m}(\Psi)$. A number of measures of $Q^{\vdash n}_{NM}$ on these bases will specify the big label $R\vdash n$, and a number of measures  $Q^{\vdash\vdash  n}_{NM}$ will specify $\mu$. \\
In \cite{BCS}, for the half-BPS sector, they define a collection of charges (momenta) whose measures completely specify $R\vdash n$, i.e. all half-BPS states. Although we haven't done it explicitly in this paper, it is clear that those momenta can be put in terms of   $Q^{\vdash n}_{NM}$ for a collection of $M$'s. In \cite{BCS} they write these momenta in terms of the Hamiltonian of $N$ fermions in a harmonic potential and use this definition to find the interpretation of the label $R$ of states in the CFT as asymptotic multipole moments of the LLM geometries in the gravity side. \\
The geometries associated with restricted Schur polynomial states $\chi_{R,\mu,m}(\Psi)$ are expected to carry the information of labels $\{R,\mu,m\}$ through asymptotic multipole moments as well. In this generic case we expect charges $\{Q^{\vdash n}_{NM},Q^{\vdash \vdash n}_{NM},Q^m_{NM}\}$ in the CFT to be the ``momenta'' which are dual to such asymptotic multipoles in the gravity solutions.
However, we cannot make the precise connection in the gravity side for generic operators. There are two obvious reasons:
\begin{itemize}
\item The eigenvalue description for generic operators is not known. So, it is impossible at this stage to make a connection with free fermions as usually done in the half-BPS case.
\item The analogs of LLM geometries for non half-BPS sectors are also unknown.
\end{itemize}
It may be instructive, and we leave it for a future work, to work out the problem in the opposite direction. We mean, trying first to find the set of solutions in SUGRA which admit a multipole expansion on labels $\{R,\mu,m\}$.\\

\paragraph{Probabilistic interpretation of the eigenvalues.}
Another suggesting point about charges $\{Q^{\vdash n}_{NM},Q^{\vdash \vdash n}_{NM},Q^m_{NM}\}$ is related to their eigenvalues. There is a nice group-theoretical interpretation of the eigenvalues of $\{Q^{\vdash n}_{NM}\}$ (and likely of the other set of charges as well) as probabilities. Remember that
\begin{equation}\label{eigenofQn}
Q^{\vdash n}_{NM}\chi_{R,\mu,m}(\Psi)=\frac{f_R^{U(N)}}{f_R^{U(M)}}\chi_{R,\mu,m}(\Psi)
\end{equation}
Take the branching graph of unitary groups. It is a graded graph whose levels are labeled by $N$. So, at level $N$ we write all the irreps of $U(N)$, and we link irreps of consecutive levels if the signatures interlace, as usual. Choosing a path in the graph from irrep $R$ at level $N$ all the way down to level 0 is tantamount to writing a Gelfand-Tselyn pattern with signature $R$. The dimension of the irrep is the number of paths we can write this way. There is a natural probability associated with this graph\footnote{Actually, equation (\ref{prob}) is a natural probability associated with any graded graph \cite{BO}, when we define the dimension of a vertex (here an irrep) as the number of paths we find all the way down.}
\begin{equation}\label{prob}
P(R,N;S,M)=\frac{\text{Dim}(R,N)}{\text{Dim}(S,M)}\text{Dim}(R,N;S,M),
\end{equation}
where $\text{Dim}(R,N;S,M)$ is the number of partial paths that start at level $M$ with irrep $S$ and end in irrep $R$ at level $N$, or equivalently, the number of partial  Gelfand-Tselyn patterns of $U(M)$ that have signature $S$ and end with signature $R$ at level $N$. It is clear that 
\begin{equation*}
\sum_{R\text{ irreps of } U(N)}P(R,N;S,M)=1.
\end{equation*}
On the one hand, it is known that
\begin{equation*}
\frac{\text{Dim}(R,N)}{\text{Dim}(S,M)}=\frac{f_R^{U(N)}}{f_S^{U(M)}}
\end{equation*}
and on the other it is not hard to see that
\begin{equation*}
\text{Dim}(R,N;R,M)=1,
\end{equation*}
that is, there is only one partial path joining irrep $R$ at level $M$ with irrep $R$ at level $N$. Thus, the eigenvalues $\frac{f_R^{U(N)}}{f_R^{U(M)}}$ of $Q^{\vdash n}_{NM}$ as shown in (\ref{eigenofQn}), are actually the probabilities of starting with irrep $R$ at level $M$ and arriving at irrep $R$ at level $N$ by a Markov process in which we take a choice of link  with probability $\frac{\text{Dim}[S,N+1]}{\text{Dim}[T,N]}$, if $S$ and $T$ are linked, at each step down.\\
Similar probabilistic interpretations are expected for the other gauge groups. Also, by combining branching graphs of the gauge groups with the branching graph of the symmetric group it should be possible to give a probabilistic interpretation to the eigenvalues of the set of charges $Q^{\vdash \vdash n}_{NM}$. It would be interesting to find out what kind of processes are the duals in the gravity side.

\section{$Q^{\vdash n}_{NM}$ acting on fermionic fields}\label{fermions}

It is true that we claim that charges $Q^{\vdash n}_{NM}$ are self-adjoint with respect to the free-field two-point function and they have eigenvectors and eigenvalues as in (\ref{eigenvectors}) but, for simplicity, we only prove it in this paper for bosonic fields. This section aims to fill this gap. It also may be taken as a warm up of the formalism.\\  
We will make it simple by considering operators in $U(N)$ theory built on one kind of  fermionic field $\psi$. Since fermionic fields are Grassmann valued we have
\begin{equation*}
\psi_{ij}\psi_{kl}=-\psi_{kl}\psi_{ij},
\end{equation*}
which introduces a modification with respect to the $1/2$-BPS sector $\Psi(Z)$ in identities like
\begin{equation}\label{fermionanticommuting}
\Psi^{\alpha(I)}_{\alpha(J)}(Z)=\Psi^I_J(Z) \longrightarrow \Psi^{\alpha(I)}_{\alpha(J)}(\psi)=\Psi^I_J(\psi) \text{sgn}(\alpha).
\end{equation}
We are going to check that $Q^{\vdash n}_{NM}=\text{Proj}_{NM}\circ \text{Av}_{MN}$ are self-adjoint and prove that the orthogonal basis found in \cite{DDN} are its eigenvectors.\\
The adjoint action acts in fermionic multitrace monomials as it did for $\frac{1}{2}$-BPS multitraces, that is,
\begin{equation*}
\text{Ad}_g (\text{Tr}_{U(N)}(\sigma \Psi))= g^{i_1}_{j_1}\psi^{j_1}_{j'_1}(g^\dagger)^{j'_1}_{i_{\sigma(1)}}\cdots g^{i_n}_{j_n}\psi^{j_n}_{j'_n}(g^\dagger)^{j'_n}_{i_{\sigma(n)}}.
\end{equation*}
Remember that since $\psi\in \mathfrak{u}(N)\hookrightarrow \mathfrak{u}(M)$, indeces $j,j'=1,\dots,M$, whereas $i=1,\dots,N$.
The averaging operator acts on multitraces as
\begin{eqnarray*}
&&\text{Av}_{_{MN}}(\text{Tr}_{U(N)}(\sigma \Psi))\nonumber \\
&=&\sum_{\alpha, \beta \in S_n}(\alpha)^I_{I'}(\sigma)_I^{I'}(\beta)^{J'}_J\Psi^J_{J'}\text{Wg}^{U(M)}(\alpha\beta)\nonumber \\
&=&\sum_{\alpha, \beta \in S_n}\text{Tr}_N(\sigma\alpha)\text{Tr}_{U(M)}(\beta \Psi)\text{sgn}(\beta)\text{Wg}^{U(M)}(\alpha\beta).
\end{eqnarray*}
In the last equality we have used (\ref{fermionanticommuting}).  We may absorb $\text{sgn}(\beta)$ into $\text{Wg}^{U(M)}(\alpha\beta)$ by
\begin{equation*}
\text{sgn}(\beta)\text{Wg}^{U(M)}(\alpha\beta)=\frac{1}{|S_n|}\sum_{R \vdash n}\frac{d_R}{f_R^{U(M)}}\chi_R(\alpha\beta)\text{sgn}(\beta)=\frac{1}{|S_n|}\sum_{R \vdash n}\frac{d_R}{f_R^{U(M)}}\chi_{R'}(\alpha\beta)\text{sgn}(\alpha),
\end{equation*}
where we have used the fact that $\chi_R(\sigma)\text{sgn}(\sigma)=\chi_{R'}(\sigma)$.\\
We may as well expand $\text{Tr}_{U(N)}(\sigma\alpha)$ in terms of characters as
\begin{equation*}
\text{Tr}_{U(N)}(\sigma\alpha)=\frac{1}{|S_n|}\sum_{S\vdash n}d_S f_S^{U(N)}\chi_S(\sigma\alpha),
\end{equation*}
and compute
\begin{eqnarray}\label{projavovertrace}
&&Q^{\vdash n}_{NM}[\text{Tr}_{U(N)}(\sigma \Psi)]\nonumber \\
&=&\frac{1}{|S_n|^2}\sum_{\alpha, \beta \in S_n}\sum_{R,S\vdash n}d_R d_S \frac{f_R^{U(N)}}{f_S^{U(M)}}\chi_{S'}(\alpha\beta)\chi_R(\sigma\alpha)\text{sgn}(\alpha)\text{Tr}_{U(N)}(\beta \Psi)\nonumber \\
&=&\frac{1}{|S_n|^2}\sum_{\alpha, \beta \in S_n}\sum_{R,S\vdash n}d_R d_S \frac{f_R^{U(N)}}{f_S^{U(M)}}\chi_{S'}(\alpha\beta)\chi_{R'}(\sigma\alpha)\text{sgn}(\sigma)\text{Tr}_{U(N)}(\beta \Psi)\nonumber \\
&=&\frac{1}{|S_n|}\sum_{\beta \in S_n}\sum_{R\vdash n} d_R \frac{f_R^{U(N)}}{f_R^{U(M)}}\chi_{R'}(\sigma\beta^{-1})\text{sgn}(\sigma)\text{Tr}_{U(N)}(\beta \Psi).
\end{eqnarray}
Now,
\begin{eqnarray*}
&&\langle Q^{\vdash n}_{NM}[\text{Tr}_{U(N)}(\sigma \Psi)] \text{Tr}_{U(N)}(\tau \bar{\Psi})\rangle\nonumber \\
&=&\frac{1}{|S_n|}\sum_{\beta \in S_n}\sum_{R\vdash n} d_R \frac{f_R^{U(N)}}{f_R^{U(M)}}\chi_{R'}(\sigma\beta^{-1})\text{sgn}(\sigma)\langle\text{Tr}_{(UN)}(\beta \Psi)\text{Tr}_{U(N)}(\tau \bar{\Psi})\rangle\nonumber \\
&=&\frac{1}{|S_n|}\sum_{\rho \beta \in S_n}\sum_{R\vdash n} d_R \frac{f_R^{U(N)}}{f_R^{U(M)}}\chi_{R'}(\sigma\beta^{-1})\text{sgn}(\sigma)\text{Tr}_{U(N)}(\beta \rho \tau \rho^{-1})\text{sgn}(\rho)\nonumber \\
&=&\frac{1}{|S_n|^2}\sum_{\rho \beta \in S_n}\sum_{R,S\vdash n} d_R d_S\frac{f_R^{U(N)}f_S^{U(N)}}{f_R^{U(M)}}\chi_{R'}(\sigma\beta^{-1})\chi_S(\beta \rho \tau \rho^{-1})\text{sgn}(\rho\sigma)\nonumber \\
&=&\frac{1}{|S_n|}\sum_{\rho \in S_n}\sum_{R \vdash n} d_R \frac{f_R^{U(N)}f_{R'}^{U(N)}}{f_R^{U(M)}}\chi_{R'}(\sigma \rho \tau \rho^{-1})\text{sgn}(\rho\sigma)\nonumber \\
\end{eqnarray*}

Consider the piece
\begin{equation}\label{selfadjointfunction}
\sum_{\rho \in S_n}\chi_{R'}(\sigma \rho \tau \rho^{-1})\text{sgn}(\sigma\rho)=\sum_{\rho \in S_n}\chi_{R'}(\sigma \rho \tau \rho^{-1})\text{sgn}(\rho),
\end{equation}
where we have made the substitution $\sigma\rho\to \rho$. It is clear that $\text{sgn}(\sigma)=1$ otherwise (\ref{selfadjointfunction}) is\footnote{Actually, we know\cite{DDN} that $\text{sgn}(\sigma)=1$ for fermions, otherwise $\text{Tr}_{U(N)}(\sigma\Psi)\equiv 0$.} 0. Thus,
\begin{equation}\label{selfadjointQnfermions}
\langle Q^{\vdash n}_{NM}[\text{Tr}_{(UN)}(\sigma \Psi)] \text{Tr}_{U(N)}(\tau \bar{\Psi})\rangle=\frac{1}{|S_n|}\sum_{\rho \in S_n}\sum_{R \vdash n} d_R \frac{f_R^{U(N)}f_{R'}^{U(N)}}{f_R^{U(M)}}\chi_{R'}(\sigma \rho \tau \rho^{-1})\text{sgn}(\rho).
\end{equation}
The RHS of (\ref{selfadjointQnfermions}) is invariant under the exchange $\sigma\leftrightarrow \tau$. For this reason we conclude that $Q^{\vdash n}_{NM}$ is also self-adjoint in the fermionic sector. \\

For the eigenvectors, we know that the restricted characters in this case is a basis of functions that have the property
\begin{equation}\label{propertyf}
 f(\rho\sigma\rho^{-1})=f(\sigma)\text{sgn}(\rho), \quad \sigma, \rho\vdash n.
\end{equation}
A basis of functions with property (\ref{propertyf}) was found in \cite{DDN}. They are labeled by self-conjugate Young diagrams, that is, diagrams which are invariant under the exchange of rows and columns, and they have the explicit expression
\begin{equation}\label{fbasis}
f_R(\sigma)=\text{Tr}\big(O_R\Gamma_R(\sigma)\big), \quad R=R' \vdash n,
\end{equation}  
where matrices $O_R$ are involutions in the carrier space of $R$ that exist only for self-conjugate representations and have the properties
\begin{equation}\label{propertiesO}
O_R=O^\dagger_R,\quad O_R O_R={\bf 1}_R, \quad O_R \Gamma_R(\sigma)=\Gamma_R(\sigma)O_R \text{sgn}(\sigma),\quad R=R'.
\end{equation}
It is straightforward to prove that a operators driven by the basis (\ref{fbasis}), which are
\begin{equation*}
\chi^{U(N)}_R(\Psi)=\sum_{\sigma\in S_n}\text{Tr}\big(O_R \Gamma_R(\sigma)\big)\text{Tr}_N(\sigma \Psi),\quad R\vdash n,
\end{equation*}
 are eigenvectors of $Q^{\vdash n}_{NM}$. Using (\ref{projavovertrace}) we have
\begin{eqnarray*}
&& Q^{\vdash n}_{NM}\bigg[\sum_{\sigma\in S_n}\text{Tr}\big(O_R \Gamma_R(\sigma)\big)\text{Tr}_{U(N)}(\sigma\Psi)\bigg] \nonumber \\
&=&\frac{1}{|S_n|}\sum_{\beta, \sigma \in S_n}\sum_{S=S'} d_S \frac{f_S^{U(N)}}{f_S^{U(M)}}\text{Tr}\big(O_R \Gamma_R(\sigma)\big)\chi_{S}(\sigma\beta^{-1})\text{sgn}(\sigma)\text{Tr}_{U(N)}(\beta \Psi)\nonumber \\
&=&\frac{1}{|S_n|}\sum_{\beta, \sigma \in S_n}\sum_{S=S'} d_S \frac{f_S^{U(N)}}{f_S^{U(M)}}\text{Tr}\big(O_R \Gamma_R(\sigma)\big)\chi_{S}(\sigma^{-1}\beta)\text{Tr}_{U(N)}(\beta \Psi)\nonumber \\
&=& \frac{f_R^{U(N)}}{f_R^{U(M)}}\sum_{\beta \in S_n} \text{Tr}\big(O_R \Gamma_R(\beta)\big)\text{Tr}_{U(N)}(\beta \Psi),
\end{eqnarray*}
where in the third line we have applied the invariance of characters under inversion and $\text{sgn}(\sigma)=1$, otherwise $\text{Tr}\big(O_R \Gamma_R(\sigma)\big)=0$.\\
Following the reasoning so far, we conclude that since $Q^{\vdash n}_{NM}$ are self-adjoint and because their eigenvalues are all different, their eigenvectors must form an orthogonal basis under the free field two-point function. Again, we can obtain the polynomial value in $N$ of the two-point function in this basis:
\begin{equation*}
\langle \sum_{\sigma\in S_n}\text{Tr}\big(O_R \Gamma_R(\sigma)\big)\text{Tr}_{U(N)}(\sigma\Psi) \sum_{\sigma'\in S_n}\text{Tr}\big(O_S \Gamma_S(\sigma')\big)\text{Tr}_{U(N)}(\sigma'\bar{\Psi})=\delta_{RS}c(R)f_R^{U(N)},
\end{equation*}    
where $c(R)$ is a number which can depend, in principle, on the diagram $R$ but not in $N$.\\

\section{Detailed construction of $Q^{\vdash\vdash n}_{NM}$}\label{Qnn}
Charges $Q^{\vdash n}_{NM}$ do not resolve the small labels of restricted Schur polynomials as can be seen in equation (\ref{cRortho}), where there is no orthogonal relation in the labels $\mu$ and $\nu$. It is reasonable though. The small labels of restricted Schur polynomials are related to the $\lambda$-structure of $\Psi$ (they do not appear in the half-BPS case, for instance).  Operators $Q^{\vdash n}_{NM}$ contain information on the total number of fields but they do not make any difference on the precise $\lambda$-structure of $\Psi$. So, why should $Q^{\vdash n}_{NM}$ care about the small labels? \\
In this section we construct charges $Q^{\vdash\vdash n}_{NM}$ that do resolve the small labels of restricted Schur polynomials. They are natural partners of $Q^{\vdash n}_{NM}$, in the sense that their construction is also dictated by the embedding.\\
It is clear that the charges we are looking for in order to resolve the small labels of restricted Schurs have to be sensitive to
the number $n_i$ of fields $\phi_i$ which  build multitrace monomials. It is reasonable that such operators carry, in principle, the label $\lambda\vdash n$ which encodes the distribution of fields inside multitrace monomials. As we are going to exploit again the embedding $G(N) \to G(M)$, we will call them $Q^\lambda_{NM}$. Later, we will see that the properties of $Q^\lambda_{NM}$ allows us to sum over all $\lambda\vdash n$ to obtain the charges
\begin{equation}\label{Q++def}
Q^{\vdash\vdash n}_{NM}=\sum_{\substack{\lambda\vdash n\\
\lambda \neq (n)}}Q^\lambda_{NM},
\end{equation}
which act non-trivially on all multitrace monomials built on $n$ fields. Note that we have substracted the partition $\lambda=(n)$ from the sum (\ref{Q++def}). Actually, $Q^{(n)}_{NM}=Q^{\vdash n}_{NM}$, so its is reasonable not to include it in the definition. As we saw in section \ref{Qn} charges  $Q^{\vdash n}_{NM}$ act non-trivially on any gauge invariant operator. We will see in this section, after constructing $Q^\lambda_{NM}$, that the charges $Q^{\vdash\vdash n}_{NM}$ defined as the sum  (\ref{Q++def}), will act non-trivially on any gauge invariant operator except for half-BPS ones, where it is 0. This is reasonable since half-BPS operators get completely fixed by labels $R\vdash n$ or, in  other words, by measures with charges $Q^{\vdash n}_{NM}$.\\
In the same spirit as for $Q^{\vdash n}_{NM}$ we are looking for
\begin{equation*}
Q^\lambda_{NM}\equiv\text{Proj}_{NM}\circ \text{Av}^\lambda_{MN}
\end{equation*}
that are self-adjoint under the free-field two-point function.
Operators $Q^\lambda_{NM}$, are not going to distinguish among the species we choose for a specific multitrace monomial. Thus, $\lambda=(n_1,\dots,n_r)$ defines completely $Q^\lambda_{NM}$ whether $n_i$ refers to bosons or fermions of any kind.\\

\subsection{$\lambda$-adjoint action}
The first step is to find an appropriate adjoint action, that we will call $\lambda$-adjoint action. Remember the compact notation we
use for fields
\begin{equation*}
\Psi\equiv\phi^{\otimes n_1}_1 \otimes \cdots \otimes \phi^{\otimes n_r}_r,
\end{equation*}
where we will use a string of indeces $I,J$
\begin{eqnarray*}
\Psi^I_J&\equiv&\Psi^{i_1}_{j_1}\Psi^{i_2}_{j_2}\cdots \Psi^{i_n}_{j_n},\qquad\qquad  \text{unitary groups}\nonumber \\
\Psi^I &\equiv& \Psi^{i_1i_2}\Psi^{i_3i_4}\cdots \Psi^{i_{2n-1}i_{2n}}, \quad \text{orthogonal and symplectic groups}.
\end{eqnarray*}
The first $n_1$ indeces of $I,J$ refer to fields $\phi_1$ and so on.\\
The multitrace monomials are encoded in permutations as
\begin{eqnarray}\label{tracesofclassicalgroups}
\text{Tr}_{U(N)}(\sigma \Psi)&=& \Psi^{i_1}_{i_{\sigma(1)}}\Psi^{i_2}_{i_{\sigma(2)}}\cdots \Psi^{i_n}_{i_{\sigma(n)}}=\Psi^I_{\sigma(I)},\quad \sigma\in S_n \nonumber \\
\text{Tr}_{SO(N)}(\sigma \Psi)&=& \Psi^{i_1i_2}\Psi^{i_3i_4}\cdots \Psi^{i_{2n-1}i_{2n}}\delta_{\sigma(I)}=\Psi^I\delta_{\sigma(I)},\quad \sigma\in S_{2n},\nonumber \\
\text{Tr}_{Sp(N)}(\sigma \Psi)&=& (J\Psi)^{i_1i_2}(J\Psi)^{i_3i_4}\cdots (J\Psi)^{i_{2n-1}i_{2n}}J_{\sigma(I)}=(J\Psi)^IJ_{\sigma(I)},\quad \sigma\in S_{2n},
\end{eqnarray}
where $J\Psi=(J\phi_1)^{\otimes n_1}\otimes \cdots \otimes (J\phi_r)^{\otimes n_r}$.\\
We will use a similar notation for the tensor product of matrices of the gauge group:
\begin{eqnarray*}
\big[g\big]&=&g_1^{\otimes n_1}\otimes g_2^{\otimes n_2}\otimes\cdots \otimes g_r^{\otimes n_r},\quad g_i \in U(N),\nonumber \\
\big[g\big]&=&g_1^{\otimes 2n_1}\otimes g_2^{\otimes 2n_2}\otimes\cdots \otimes g_r^{\otimes 2n_r},\quad g_i \in SO(N), Sp(N)
\end{eqnarray*} 
or explicitely
\begin{eqnarray*}
\big[g\big]^I_J&\equiv&(g_1)^{i_1}_{j_1}\cdots (g_1)^{i_{n_1}}_{j_{n_1}}(g_2)^{i_{n_1+1}}_{j_{n_1+1}}\cdots (g_2)^{i_{n_1+n_2}}_{j_{n_1+n_2}}\cdots (g_r)^{i_{n-n_r+1}}_{j_{n-n_r+1}}\cdots (g_r)^{i_n}_{j_n}\nonumber \\
\big[g\big]_{IJ} &\equiv& (g_1)_{i_1j_1}\cdots (g_1)_{i_{2n_1}j_{2n_1}}\cdots\cdots (g_r)_{i_{2n-2n_r+1}j_{2n-2n_r+1}}\cdots (g_r)_{i_{2n}j_{2n}}.
\end{eqnarray*}
Let us first consider the unitary case. 
The first (naive) approach to the $\lambda$-adjoint action would be
to define $\widehat{\text{Ad}}^{\lambda}_{[g]}$ which acts on GI operators in a way that the adjoint action on $\phi_i$ fields is $g_i\phi_ig_i^{-1}$. This adjoint action clearly distinguishes between $r$ different fields. 
Let us see how $\widehat{\text{Ad}}^{\lambda}_{[g]}$ acts on multitrace monomials. In the case of unitary groups
\begin{eqnarray}\label{naivead}
&&\widehat{\text{Ad}}^{\lambda}_{[g]}[\text{Tr}_{U(N)}(\sigma \Psi] \nonumber \\
&=& (\sigma)^{I'}_I\bigg[(g_1)^{i_1}_{j_1}(\phi_1)^{j_1}_{j'_1}(\bar{g}_1)^{i'_1}_{j'_1}\cdots (g_1)^{i_{n_1}}_{j_{n_1}}(\phi_1)^{j_{n_1}}_{j'_{n_1}}(\bar{g}_1)^{i'_{n_1}}_{j'_{n_1}}~ (g_2)^{i_{n_1+1}}_{j_{n_1+1}}(\phi_2)^{j_{n_1+1}}_{j'_{n_1+1}}(\bar{g}_2)^{i'_{n_1+1}}_{j'_{n_1+1}}\cdots \bigg]\nonumber \\
&=&\big[g\big]^I_J \big[\bar{g}\big]^{I'}_{J'} \big[\Psi\big]^J_{J'} (\sigma)^{I'}_I,
\end{eqnarray}
where indices $I$ run from 1 to $N$ and indeces $J$ from 1 to $M$. The naive averaging
operator would be:
\begin{equation}\label{naiveav}
\int_{g_1,\dots,g_r\in U(M)}\text{d}\big[g\big] ~ \widehat{\text{Ad}}^{\lambda}_{[g]}[\mathcal{O}(\Psi)],
\end{equation}
where $\text{d}\big[g\big]=\text{d}g_1\cdots \text{d}g_r$. Operators (\ref{naiveav}) are maps between GI in $U(N)$ and GI operators in $U(M)$, as expected, since the integral contracts indeces $J$ with $J'$ in (\ref{naivead}). However, one can see that the result of aplying   (\ref{naiveav}) on multitrace monomials is a    linear combination of
operators of the type  $\text{Tr}_{U(M)}(\rho \Psi)$, where $\rho\in S_\lambda$.  Multitrace monomials of this
kind do not mix fields $\phi_i$ and $\phi_{i'}$ in the same trace, so there is no hope that generic operators that involve traces like, say,
$\text{Tr}(\phi_1\phi_2)$ could be eigenvectors of (\ref{naiveav}) after projection $\text{Proj}_{NM}$.  \\  

The problem resides in the adjoint action (\ref{naivead}). As we can see in  (\ref{naivead}), it involves $n$ adjoint actions distributed as $n_i$ $g_i$-adjoint actions for $i=1,\dots,r$. This is the right spirit. However, the group element $g_i$ acts only on $\phi_i$'s. This is the reason for not mixing $\phi_i$'s and $\phi_{i'}$'s in the same trace of the operators we get under the averaging action. The problem gets solved if we allow one of the two matrices of the single adjoint actions to act on an arbitrary slot. In order to do it democratically, we shuffle all the left hand side matrices\footnote{Equivalently we could shuffle the right hand side matrices. See Appendix \ref{GAA}.}   of the adjoint actions over the fields and sum over all possible shufflings. Let us see how to do it.\\
The idea is to replace
\begin{equation}\label{adjointactiononfields}
\widehat{\text{Ad}}^{\lambda}_{[g]}(\Psi)=\big[g\big]^I_J \big[\bar{g}\big]^{I'}_{J'} \big[\Psi\big]^J_{J'} \to \text{Ad}^{\lambda}_{[g]}(\Psi) =\frac{1}{|S_n|}\sum_{\alpha\in S_n}\big[g\big]^I_{\alpha(J)} \big[\bar{g}\big]^{I'}_{J'} \big[\Psi\big]^J_{J'}.
\end{equation}
But one must be careful in doing so. We should permute the slots where $g_i$ but keeping the structure of the multitrace monomials, which is carried by indeces $I$. In other words, we should write
\begin{equation*}
\text{Ad}^{\lambda}_{[g]}\text{Tr}_{U(N)}(\sigma\Psi) =\frac{1}{|S_n|}\sum_{\alpha\in S_n}\big[g\big]^I_{\alpha(J)} \big[\bar{g}\big]^{I'}_{J'} \big[\Psi\big]^J_{J'}(\sigma')^{I'}_I,
\end{equation*}
for some $\sigma'\in S_n$ which is related to $\sigma$ and $\alpha$. A careful analysis on indeces reveals that $\sigma'=\alpha^{-1}\sigma$,  see Appendix \ref{GAA} for details.  All in all, the correct adjoint action on the operators must be defined as:
\begin{equation}\label{correctad}
\text{Ad}^{\lambda}_{[g]}[\text{Tr}_{U(N)}(\sigma \Psi)]=\frac{1}{|S_n|}\sum_{\alpha \in S_n} \big[g\big]^I_{\alpha(J)} \big[\Psi\big]^J_{J'} \big[\bar{g}\big]^{I'}_{J'}(\alpha^{-1}\sigma)^{I'}_I.
\end{equation}

Some comments about (\ref{correctad}) are in order.
\begin{itemize}
\item After the shuffling some of the fields will be acted on as $g_{i'}\phi_i\bar{g}_i$, that is, $\text{Ad}^{\lambda}_{[g]}$ is {\it not} a collection of truly adjoint actions on every field as  $\text{Ad}_g$ is. With this fact in mind we keep on calling it $\lambda$-adjoint action.
\item We see that definition (\ref{correctad}) reduces to $\text{Ad}_g$ when $\lambda=(n)$, that is, when the multitrace are built on just one letter of the alphabet. In that case, the action of shuffling is trivial so it can be omitted.
\item There is a conceptual meaning about shuffling the group elements $g_i$ in the $\lambda$-adjoint action. If we compare (\ref{correctad}) to the naive approach (\ref{naivead}) we see that in (\ref{correctad}) $g_i$'s are  no longer associated to
fields $\phi_i$'s. The delocalization of $g_i$'s makes them refer to the number of fields $n_i$, instead of the fields $\phi_i$ themselves.
\item We have decided to `shuffle-act' on the left, that is moving unbarred elements $g_i$. We could have acted on the right by moving barred elements with identical result. Both actions commute. However, applying both at the same time spoils the properties of the operator. There is a freedom, however, of fully acting on one side (as we have done) and acting on the other with an appropriate subgroup of $S_n$, perhaps abelian. This leaves a room for constructing $Q_{NM}^m$ which will commute with $Q^{\vdash n}_{NM}$ and with $Q^{\vdash\vdash n}_{NM}$ and will resolve the multiplicities, which are not resolved by $Q^{\vdash\vdash n}_{NM}$ with (\ref{correctad}) as we will see later. We will investigate and report $Q_{NM}^m$ elsewhere.
\item We claim that the $\lambda$-adjoint action (\ref{correctad}) is valid for generic operators built on a distribution $\lambda=(n_1,\dots,n_r)$ of fields and also for any classical gauge group with minor modifications as we are going to see. What we mean with this is that (\ref{correctad}) can be used to construct operators $Q^\lambda_{NM}=\text{Proj}_{NM}\circ\text{Av}_{MN}^{\lambda}$ which will be self-adjoint with respect to the free field two-point function and whose eigenvectors are restricted Schur polynomials. Moreover, the action of $Q^\lambda_{NM}$ together with
$Q^{\vdash n}_{NM}=\text{Proj}_{NM}\circ\text{Av}_{MN}$ will serve to obtain the value (up to constants) of the free correlators of generic operators.
\end{itemize} 
For the orthogonal gauge group we have
\begin{equation*}
\text{Ad}^{\lambda}_{MN}\big(\text{Tr}_{SO(N)}(\sigma\Psi)\big)=\frac{1}{|S_{2n}|}\sum_{\alpha\in S_{2n}}\big[g\big]_{I\alpha(J)}\Psi^J \delta_{\alpha^{-1}\sigma(I)},\quad g\in SO(M),
\end{equation*}
where we have made the same reasoning as in the unitary case for the change $\sigma\to\alpha^{-1}\sigma$.\\
For the symplectic case we have
\begin{equation*}
\text{Ad}^{\lambda}_{MN}\big(\text{Tr}_{Sp(N)}(\sigma\Psi)\big)=\frac{1}{|S_{2n}|}\sum_{\alpha\in S_{2n}}\big[g\big]_{I\alpha(K)}(J\Psi)^K J_{\alpha^{-1}\sigma(I)},\quad g\in Sp(M).
\end{equation*}

\subsection{Properties of $Q^{\vdash \vdash n}_{NM}$}\label{propertiesQnn}
Now, with the adjoint actions so-defined in the last subsection, we define the $\lambda$-averaging operators as 
\begin{equation}\label{goodav}
\text{Av}^\lambda_{MN}[\mathcal{O}(\Psi)]\equiv\int_{g_1,\dots,g_r\in G(M)}\text{d}\big[g\big] ~ \text{Ad}^{\lambda}_{[g]}[\mathcal{O}(\Psi)].
\end{equation}
Now, we compose it with projections and define
 \begin{equation*}
Q^\lambda_{NM}\equiv\text{Proj}_{NM}\circ\text{Av}^\lambda_{MN},
\end{equation*}
which will be maps of gauge invariant operators built on $\phi_i\in \mathfrak{g}(N)$. Let us summarize the properties of these charges. Detailed proofs are postponed to section \ref{proofQnn}.
\begin{itemize}
\item $Q^\lambda_{NM}$ are self-adjoint under the free-field two point function of the theory.
\item They only depend on $\lambda=(n_1,\dots,n_r)$, and not on the fermionic or bosonic nature of the  fields.
\item Their eigenvectors are restricted Schur polynomials:
 \begin{eqnarray}\label{Qlambdas}
Q_{NM}^\lambda(\chi^{U(N)}_{R,\mu,ij}(\Psi))&=&\frac{|S_\lambda|}{|S_n|}\frac{f_R^{U(N)}}{f_\mu^{U(M)}}\chi^{U(N)}_{R,\mu,ij}(\Psi) \nonumber \\
Q_{NM}^\lambda(\chi^{SO(N)}_{R,\mu,i}(\Psi))&=&\frac{|S_\lambda|}{|S_\lambda[S_2]|^2|S_{2n}|}\frac{f_R^{SO(N)}}{f_\mu^{SO(M)}}\chi^{SO(N)}_{R,\mu,i}(\Psi) \nonumber \\
Q_{NM}^\lambda(\chi^{Sp(N)}_{R,\mu,i}(\Psi))&=&\frac{|S_\lambda|}{|S_\lambda[S_2]|^2|S_{2n}|}\frac{f_R^{Sp(N)}}{f_\mu^{Sp(M)}}\chi^{Sp(N)}_{R,\mu,i}(\Psi),
\end{eqnarray}
\item Since their eigenvalues are all different for each $\mu$ irrep of $S_\lambda$ and they are self-adjoint, the two point function is forced to be orthogonal in the labels $\mu$. This is the resolution of small labels we mentioned above. If we link this fact with the correlator form  (\ref{cRortho}), we can see that by means of $\{Q^{\vdash n}_{NM}\}$ and $\{Q_{NM}^\lambda\}$ and, ultimately, by means of the embedding structure, the form of the correlators is necessarily
 \begin{equation*}
\langle\chi^{G(N)}_{R,\mu,m},\chi^{G(N)}_{S,\nu,m'}\rangle=c(R,\mu,m,m')f_R^{G(N)}\delta_{RS}\delta_{\mu\nu},
\end{equation*}
where the polynomials $f_R^{G(N)}$ are naturally found for each gauge group.
\item If we consider $\Psi$ with a given $\lambda'$-structure we can see that
\begin{equation*}
Q_{NM}^\lambda[\mathcal{O}(\Psi)]\equiv 0,
\end{equation*}
for $\lambda\neq \lambda'$. So, it is natural to define
\begin{equation*}
Q_{NM}^{\vdash\vdash n}\equiv \sum_{\substack{\lambda\vdash n\\
\lambda \neq (n)}}Q_{NM}^\lambda,
\end{equation*}
which, alike $Q_{NM}^{\vdash n}$, will act non-trivially on all gauge invariant operators (except for half-BPS operators, for which the action of $Q_{NM}^{\vdash\vdash n}$ is 0) built on a total number of $n$ fields. 
\end{itemize}

\section{Characterization of restricted character bases via the convolution product}\label{CRBCP}
One of the handicaps we face in this work is that the restricted character basis for orthogonal and symplectic groups is still under development (the $\frac{1}{2}$-BPS sector has been worked out\cite{CDD,CDD2} and, recently the 1/4-BPS sector for orthogonal groups\cite{GK1,GK2}), so in order to claim that the charges $Q^{\vdash n}_{NM}$ and $Q^{\vdash\vdash n}_{NM}$ we have built actually single out restricted Shcur polynomials via eigenvectors we should give a characterization of the restricted character bases in a way that can be extended to the orthogonal and symplectic cases. \\
For unitary gauge groups, the Schur polynomial basis (1/2-BPS case) corresponds to characters\cite{CJR}. Restricted Schur polynomials are driven by the restricted character basis, which has also been developed\cite{DSS,BCD}. We are going to see that the restricted character basis can be uniquely characterized by a set of convolution relations. The set is complete, in the sense that it closes an algebra under the convolution product. To include fermions in the game we need to extend the algebra of convolution. We will learn from this process to tackle the other gauge groups. \\
For any functions $g,f:S_n\to \mathbb{C}$ we define the convolution product as
\begin{equation}\label{convproduct}
f\star g~ (\sigma)\equiv \sum_{\alpha\in S_n}f(\alpha^{-1})g(\alpha\sigma).
\end{equation}
The vector space of all functions of $S_n$ on $\mathbb{C}$ will be called $\mathcal{H}$. It is clear that $\mathcal{H}$ equipped with (\ref{convproduct}) form an algebra. Product (\ref{convproduct}) corresponds to the usual product in the group algebra, that is,
\begin{equation*}
(f\star g)'=f'g',
\end{equation*}
where prime is the usual map between functions and elements of the group algebra:
\begin{equation*}
f'=\sum_{\sigma\in S_n}f(\sigma)\sigma.
\end{equation*}
The algebra of convolution of $\mathcal{H}$ has a unit, which is the function 
\begin{equation*}
\delta(\sigma)=
 \left\{\begin{array}{rl}
1 &\text{if } \sigma=\text{id},\\
0 & \text{ohtherwise},
\end{array}\right.
\end{equation*}
so
\begin{equation*}
f\star \delta =\delta\star f=f,
\end{equation*}
but is in general non-commutative because
\begin{equation*}
f\star g ~(\sigma)=\sum_{\alpha\in S_n}f(\alpha^{-1})g(\sigma\alpha)\neq \sum_{\alpha\in S_n}g(\alpha^{-1})f(\sigma\alpha)=g\star f~(\sigma).
\end{equation*}
We are interested in the subalgebras of $\mathcal{H}$ that play a role in our operators, they will be algebras of functions with a certain symmetry which is dictated by the symmetry of multitrace monomials. In the subsequent subsections we will treat them all.

\subsection{Unitary groups}
Let us consider the bosonic case first and at the end of this subsection we will see how to deal with fermions. For the unitary group, we have seen that the multitrace monomials $\text{Tr}(\sigma\Psi)$ are invariant under the change $\sigma\to \gamma\sigma\gamma^{-1}$, where $\gamma\in S_n$ for the 1/2-BPS case, and $\gamma\in S_\lambda$ if we consider operators built on multiple bosonic fields. So we will restrict ourselves to the subalgebra of $\mathcal{H}$ of functions which are constant on a given orbit  of $\sigma$ generated by $\gamma\sigma\gamma^{-1}$.

\paragraph{Half BPS functions}
For half BPS operators the Schur functions are characters. We know that for characters
\begin{equation*}
\chi_R\star \chi_S~(\sigma)=\sum_{\alpha\in S_n}\chi_R(\alpha^{-1})\chi_S(\alpha\sigma)=\delta_{RS}\frac{n!}{d_R}\chi_R(\sigma),\quad R\vdash n.
\end{equation*}
Now, for the sake of simplicity in the algebra, we will normalize characters like
\begin{equation*}
b^{U}_R(\sigma)\equiv \frac{d_R}{n!}\chi_R(\sigma).
\end{equation*}
Then, we have the relations
\begin{equation}\label{bcharactersalgebra}
b^{U}_R\star b^{U}_S=\delta_{RS}b^{U}_R, \quad R,S\vdash n.
\end{equation}
Relations (\ref{bcharactersalgebra}) completely define functions $b^{U}_R$, and so they define characters. The algebra
of class functions  has the unit $\delta\star b^{U}_R=b^{U}_R\star \delta=b^{U}_R$, which can be expanded as
\begin{equation*}
\delta(\sigma)=\sum_{R\vdash n}b^{U}_R(\sigma).
\end{equation*}

\paragraph{General bosonic functions}
When our operators are composed of more than one kind of field, say we have $n_i$ times field $\phi_i$, the multitraces have the symmetry
\begin{equation*}
\text{Tr}_{U(N)}(\gamma \sigma\gamma^{-1}\Psi)=\text{Tr}_{U(N)}(\sigma\Psi), \quad \sigma\in S_n, \quad \gamma\in S_\lambda.
\end{equation*}
So, we are interested in functions of $S_n$ that have the symmetry
\begin{equation*}
b^{U}(\gamma\sigma\gamma^{-1})=b^{U}(\sigma), \quad \sigma\in S_n, \quad \gamma\in S_\lambda.
\end{equation*}
Before defining the restricted character basis by means of their convolution relations, let us study the algebra relations of characters of $S_\lambda$.
Similar relations to (\ref{bcharactersalgebra}) are found when we consider characters of a $S_\lambda\subset S_n$.
Remember that an irrep of $S_\lambda$ is labeled by $\mu=(r_1\vdash n_1, \dots, r_{l(\lambda)}\vdash n_{l(\lambda)})$, where $n_1+n_2+\cdots +n_{l(\lambda)}=n$, and $d_\mu=d_{r_1}\cdots d_{l(\lambda)}$. Characters will be normalized
as\footnote{Note that the difference in the notation of $b^{U}_R(\sigma)$ and $b^{U}_\mu(\sigma)$ relies only in the labels. It should not lead to much confusion.} 
\begin{equation*}
b^{U}_\mu(\rho)\equiv\frac{d_\mu}{|S_{\lambda}|}\chi_\mu(\rho),\quad \rho\in S_{\lambda}, 
\end{equation*}
and the algebra of class functions of $\rho \in S_{\lambda}$ in this basis is
\begin{equation*}
 b^{U}_\mu\star  b^{U}_\nu~(\rho)=\delta_{\mu\nu} b^{U}_\mu(\rho),
\end{equation*}
with unit
\begin{equation*}
\delta(\rho)=\sum_\mu b^{U}_\mu(\rho).
\end{equation*}
$\{b^{U}_\mu\}$ are a basis of class functions of $S_{\lambda}$. We will extend these functions to apply on $\sigma\in S_n$ by the definition
\begin{equation*}
\dot{b}^{U}_\mu(\sigma)= \left\{\begin{array}{rl}
b^{U}_\mu(\sigma) &\text{if } \sigma\in S_{\lambda} ,\\
0 & \text{otherwise},
\end{array}\right.
\end{equation*}
It is easy to see that 
\begin{equation*}
\delta(\sigma)=\sum_\mu \dot{b}^{U}_\mu(\sigma), \quad \sigma\in S_n.
\end{equation*}
\\
Now we go back to the restricted character basis. As said above, it has been completely studied. They are defined as
\begin{equation*}
\chi_{R,\mu,ij}(\sigma)=\text{Tr}(P_{R\to (\mu)ij} \Gamma_R(\sigma)), \quad i,j=1,\dots,g(R;\mu),
\end{equation*} 
where $g(R;\mu)$ are the Littlewood-Richardson coefficients, that is, the number of times irrep $\mu$ of $S_{\lambda}$ appears when $R$ irrep of $S_n$ is restricted to $S_{\lambda}$. Indeces $i,j$ label the copies of $\mu$ when subduced from $R$. Matrices $P_{R\to (\mu)ij}$ act as projectors when $i=j$ from the carrier space of $R$ to the the carrier space of $\mu$ but they intertwine copies $i$ and $j$ when $i\neq j$. See more details in\cite{DSS,BCD}.  
From the Schur orthogonality of irreps
\begin{equation}\label{Schurorthogonality}
\frac{1}{|S_n|}\sum_{\sigma\in S_n}\Gamma^R_{op}(\sigma)\Gamma^S_{qr}(\sigma^{-1})=\frac{1}{d_R}\delta_{RS}\delta_{or}\delta_{pq}, \quad R,S \vdash n,
\end{equation}
and the properties of $P$:
\begin{eqnarray}\label{propertiesprojectors}
P_{R,\mu,ij}P_{S,\nu,kl}&=&\delta_{RS}\delta_{\mu\nu}\delta_{jk}P_{R,\mu,il}\nonumber \\
P_{R,\mu,ij}&=&P_{R,\mu,ij}^\dagger
\end{eqnarray}
 we can see that the convolution product of these functions is
\begin{equation}\label{restrictedschurconvolution}
\chi_{R,\mu,ij}\star \chi_{T,\nu,kl}=\delta_{RS}\delta_{\mu\nu}\delta_{jk}\frac{|S_n|}{d_R}\chi_{R,\mu,il}.
\end{equation}
Again, normalizing as
\begin{equation*}
b^{U}_{R,\mu,ij}\equiv \frac{d_R}{|S_n|}\chi_{R,\mu,ij}
\end{equation*}
we get the (non-commutative) relations
\begin{equation}\label{brestrictedschurconvolution}
b^{U}_{R,\mu,ij}\star b^{U}_{S,\nu,kl}=\delta_{RS}\delta_{\mu\nu}\delta_{jk}b^{U}_{R,\mu,il}
\end{equation}
From (\ref{brestrictedschurconvolution}) we can see that in this basis the unit may be expanded as
\begin{equation}\label{unitrestricted}
\delta(\sigma)=\sum_{R,\mu,i}b^{U}_{R,\mu,ii}(\sigma),\quad \sigma\in S_n.
\end{equation}
Relations (\ref{brestrictedschurconvolution}) completely determine the basis $\{b_{R,\mu,ij}\}$ and so the restricted character basis. However, as happens in our case, they are sometimes not useful for computations. We will need the convolution relations of restricted characters when combined with characters of both $S_n$ and $S_{\lambda}$. Using Schur orthogonality and the projector properties of $P$ we get the commutative relations
\begin{eqnarray}\label{usefulrestricted}
b^{U}_R\star b^{U}_{S,\mu,ij}&=&\delta_{RS}b^{U}_{S,\mu,ij},\nonumber \\
\dot{b}^{U}_\mu \star b^{U}_{S,\nu,ij}&=&\delta_{\mu\nu} b^{U}_{S,\nu,ij}, \nonumber \\
b^{U}_R \star \dot{b}^{U}_\mu &=&\sum_i  b^{U}_{R,\mu,ii},
\end{eqnarray}
where the third set of relations are obtained by combining the former two with (\ref{unitrestricted}). The commutativity of the second set of relations in (\ref{usefulrestricted}) can be easily seen if we take into account that $\dot{b}^{U}_\mu(\sigma)=\dot{b}^{U}_\mu (\sigma^{-1})$ for all $\sigma \in S_n$, and that $\dot{b}^{U}_\mu$ is 0 for all elements outside $S_\lambda$. Then
\begin{eqnarray*}
\dot{b}^{U}_\mu \star b^{U}_{S,\nu,ij}~(\sigma)&=&\sum_{\alpha\in S_n}\dot{b}^{U}_\mu(\alpha^{-1}) b^{U}_{S,\nu,ij}(\alpha\sigma)=\sum_{\alpha\in S_n}\dot{b}^{U}_\mu(\alpha^{-1}) b^{U}_{S,\nu,ij}(\sigma\alpha)\nonumber \\
&=&\sum_{\alpha\in S_n}\dot{b}^{U}_\mu(\alpha^{-1}\sigma) b^{U}_{S,\nu,ij}(\alpha)= b^{U}_{S,\nu,ij}\star \dot{b}^{U}_\mu~ (\sigma).
\end{eqnarray*}
 The set 
(\ref{usefulrestricted}) partially determines the restricted character basis. Namely, they determine the basis up to
multiplicities or, in other words, they determine the commutative blocks of the algebra. But this is enough for our purposes, since the charges $Q^{\vdash n}_{NM}$  that we are considering in this paper do not resolve the multiplicities.

\paragraph{Adding fermions}
If some of the fields that build our operator are fermionic the symmetries of the multitrace monomials make us consider functions
which have the property
\begin{equation}\label{fermionicfunctions}
f(\gamma\sigma\gamma^{-1})=f(\sigma)\text{sgn}(\gamma^f), \quad \gamma=\gamma_B\circ \gamma_F\in S_\lambda.
\end{equation}
Functions (\ref{fermionicfunctions}) do not close any algebra under the convolution product since, if $f$ and $g$ fulfill (\ref{fermionicfunctions}), we have
\begin{eqnarray*}
f\star g~(\sigma)&=&\sum_{\alpha\in S_n}f(\alpha^{-1})g(\alpha\sigma)=\frac{1}{|S_\lambda|}\sum_{\substack{\alpha\in S_n\\
\gamma\in S_\lambda}}f(\gamma\alpha^{-1}\gamma^{-1})g(\alpha\sigma)\text{sgn}(\gamma^f)\nonumber \\
&=&\frac{1}{|S_\lambda|}\sum_{\substack{\alpha\in S_n\\
\gamma\in S_\lambda}}f(\alpha^{-1})g(\gamma\alpha\gamma^{-1}\sigma)\text{sgn}(\gamma^f)\nonumber \\
&=&\frac{1}{|S_\lambda|}\sum_{\substack{\alpha\in S_n\\
\gamma\in S_\lambda}}f(\alpha^{-1})g(\alpha\gamma^{-1}\sigma\gamma),
\end{eqnarray*}
and the last line is manifestly invariant under $\sigma\to \gamma\sigma\gamma^{-1}$, so the product of two fermionic functions gives a bosonic function. With the same analysis we can see that the convolution product of a fermion and a boson gives a fermion. In order to close an algebra it is necessary to consider both fermionic and bosonic functions. Let us use $f$ to denote fermionic functions and $b$ to denote bosonic ones.\\
The first issue to discuss is the set of labels of $f$. When $\text{sgn}(\gamma^f)$ appears in (\ref{fermionicfunctions}) the orbits seem to split. One could think that we should consider different functions (and so different labels) for the orbits where $\gamma^f$ is even and for $\gamma^f$ odd. However, because the difference is just a sign both functions are linearly dependent, so the same label must be used for both orbits. Moreover, it is clear from (\ref{fermionicfunctions}) that if $\sigma$  commutes with any odd permutation $\gamma\in S_\lambda$, all the fermionic functions of $\sigma$ vanish. So, orbits that contain one such $\sigma$ must be excluded. At the end of the day, we are left with a collection of labels which is a subset of the labels used in the purely bosonic case. \\ 
Indeed, the `valid' labels for fermions have been found to be the ones that have self-conjugate representations (not necessarily irreducible) in the fermionic subgroup of $S_\lambda$ \cite{DDN}. It was proved in \cite{DDN} that for those labels, and only for them, one can construct an involution that take bosonic functions into fermionic ones. This involution goes schematically like
\begin{equation*}
\text{Tr}(P\Gamma(\sigma))\to \text{Tr}(OP\Gamma(\sigma)),
\end{equation*}  
where $O\Gamma(\gamma)=\text{sgn}(\gamma)\Gamma(\gamma)O$ for $\gamma\in S_{\lambda}$, with properties
\begin{equation*}
O=O^+,\quad O^2={\bf 1},\quad [O,P]=0.
\end{equation*}
Functions $\text{Tr}(OP\Gamma(\sigma))$ have the symmetry (\ref{fermionicfunctions}) and their labels are a subset of the labels for bosonic functions. Applying Schur orthogonality, properties of $P$'s and the properties of $O$ we arrive to an extension of the algebra  (\ref{brestrictedschurconvolution}) that includes fermions 
\begin{eqnarray*}
b^{U}_{R,\mu,ij}\star b^{U}_{S,\nu,kl}&=&\delta_{RS}\delta_{\mu\nu}\delta_{jk}b^{U}_{R,\mu,il}\nonumber \\
f^{U}_{R,\mu,ij}\star b^{U}_{S,\nu,kl}&=&\delta_{RS}\delta_{\mu\nu}\delta_{jk}f^{U}_{R,\mu,il}\nonumber \\
f^{U}_{R,\mu,ij}\star f^{U}_{S,\nu,kl}&=&\delta_{RS}\delta_{\mu\nu}\delta_{jk}b^{U}_{R,\mu,il},
\end{eqnarray*}
keeping in mind that $f^{U}_{R,\mu,ij}\equiv 0$ for `non-valid' labels.

\subsection{Orthogonal groups}
For orthogonal gauge groups, multitrace monomials have the symmetries
\begin{equation*}
\text{Tr}(\eta\sigma\xi\Psi)=\text{Tr}(\sigma \Psi)\text{sgn}(\eta),\quad \eta\in S_\lambda[S_2], \quad \xi\in S_n[S_2],\quad \sigma\in S_{2n}.
\end{equation*}
It is easy to see that functions 
\begin{equation*}
b^{SO}(\eta\sigma\xi)=b^{SO}(\sigma)\text{sgn}(\eta) \quad \eta\in S_\lambda[S_2], \quad \xi\in S_n[S_2],\quad \sigma\in S_{2n}
\end{equation*}
do not close any algebra under the convolution product on their own, since for any $b^{SO},b'^{SO}$ with the above properties we have
\begin{eqnarray*}
b^{SO}\star b'^{SO}~(\sigma)&=&\sum_{\alpha\in S_{2n}}b^{SO}(\alpha^{-1})b'^{SO}(\sigma\alpha)\nonumber \\
&=&\frac{1}{|S_\lambda|}\sum_{\substack{\alpha\in S_{2n}\\
\eta\in S_\lambda[S_2]}}b^{SO}(\eta\alpha^{-1})b'^{SO}(\sigma\alpha)\text{sgn}(\eta)\nonumber \\
&=&\frac{1}{|S_\lambda|}\sum_{\substack{\alpha\in S_{2n}\\
\eta\in S_\lambda[S_2]}}b^{SO}(\alpha^{-1})b'^{SO}(\sigma\alpha\eta)\text{sgn}(\eta)\nonumber \\
&=&\frac{1}{|S_\lambda|}\sum_{\substack{\alpha\in S_{2n}\\
\eta\in S_\lambda[S_2]}}b^{SO}(\alpha^{-1})b'^{SO}(\sigma\alpha)\text{sgn}(\eta)=0.
\end{eqnarray*}
Indeed, these functions will form part of a broader algebra. We will see shortly that the complete algebra involves functions of the type
\begin{eqnarray*}
b^+(\eta\sigma\xi)&=&b^+(\sigma), \quad \sigma\in S_{2n} \quad \eta, \xi\in S_n[S_2] \text{~or~} \eta, \xi\in S_\lambda[S_2] , \nonumber \\
b^{SO}(\eta\sigma\xi)&=&b^{SO}(\sigma)\text{sgn}(\eta),\quad \sigma\in S_{2n}  \quad \eta, \in S_\lambda[S_2],\quad \xi\in S_n[S_2] 
\end{eqnarray*}
The choice of $\eta, \xi\in S_n[S_2]$ or  $\eta, \xi\in S_\lambda[S_2]$ in the first definition will be clear from the labels they carry.\\
The algebra of functions $b^+$ in both, the case where the functions are invariant in the double coset $S_n[S_2]\backslash S_{2n}/S_n[S_2]$ and the case where the functions are invariant in the double coset $S_\lambda[S_2]\backslash S_{2n}/S_\lambda[S_2]$, has a unit $\delta_\lambda^+(\sigma)$ which is also invariant on the double coset\footnote{Note that $\delta(\sigma)$ is not invariant under the double coset, so it does not belong to the algebra of $b^+$.} and is defined as
\begin{equation}\label{unitbplus}
\delta_\lambda^+(\sigma)=\frac{1}{|S_\lambda[S_2]|}\sum_{\xi\in S_\lambda[S_2]}\delta(\xi \sigma),\quad \sigma \in S_{2n},
\end{equation}
which is 0 unless $\sigma\in S_\lambda[S_2]$,  in which case it equals 1.\\

The algebra that contains functions $b^{SO}$ is going to be expressed in terms of combinations $(b^+,b^{SO})$  together with $(b^+, b^+)$. Note that this algebra will be  non-commutative. For example
\begin{eqnarray*}
b^+\star b^{SO}~(\sigma)&=& \sum_{\alpha\in S_{2n}}b^+(\alpha^{-1})b^{SO}(\sigma\alpha)\nonumber \\
&=&\frac{1}{|S_n|}\sum_{\substack{\alpha\in S_{2n}\\
\eta\in S_n[S_2]}}b^+(\alpha^{-1}\eta)b^{SO}(\sigma\alpha)\nonumber \\
&=&\frac{1}{|S_n|}\sum_{\substack{\alpha\in S_{2n}\\
\eta\in S_n[S_2]}}b^+(\alpha^{-1})b^{SO}(\sigma\eta\alpha)
\end{eqnarray*}
is, in general, a nonzero function of $\sigma$ of type $b^{SO}$, whereas
\begin{eqnarray*}
b^{SO}\star b^+~(\sigma)&=& \sum_{\alpha\in S_{2n}}b^{SO}(\alpha^{-1})b^+(\sigma\alpha)\nonumber \\
&=&\frac{1}{|S_\lambda|}\sum_{\substack{\alpha\in S_{2n}\\
\eta\in S_\lambda[S_2]}}b^{SO}(\eta\alpha^{-1})b^+(\sigma\alpha)\text{sgn}(\eta) \nonumber \\
&=&\frac{1}{|S_\lambda|}\sum_{\substack{\alpha\in S_{2n}\\
\eta\in S_\lambda[S_2]}}b^{SO}(\alpha^{-1})b^+(\sigma\alpha\eta)\text{sgn}(\eta) \nonumber \\
&=&\frac{1}{|S_\lambda|}\sum_{\substack{\alpha\in S_{2n}\\
\eta\in S_\lambda[S_2]}}b^{SO}(\alpha^{-1})b^+(\sigma\alpha)\text{sgn}(\eta)=0.
\end{eqnarray*}
Now, the first thing to discuss is the labels we should use for these functions, and for that matter we are going
to reproduce a general result of finite groups coming from Mackey's theory that can be found (with its proof) in \cite{DB}.
\begin{theorem}[Geometric form of Mackey's Theorem]
Let $H_1$ and $H_2$ be subgroups of the finite group $G$, and let $\psi_i$ be a linear character of $H_i$. Let $\Lambda \in \text{Hom}_G(\psi_1^G,\psi_2^G)$. Then there exists a function $\Delta: G\to C$ such that
\begin{equation}\label{theorem}
\Delta(h_2 g h_1)=\psi_2(h_2)\Delta(g)\psi_1(h_1), \quad h_i\in H_i
\end{equation}
and $\Lambda f=\Delta\star f$ for all $f\in \psi_1^G$. The map $\Lambda\to \Delta$ is a vector space isomorphism of $\text{Hom}_G(\psi_1^G,\psi_2^G)$ with the space of all functions satisfying (\ref{theorem}).
\end{theorem}
We will make use of this results in the following paragraphs.

\paragraph{Functions $b^+$ with $\eta, \xi\in S_n[S_2]$.}
These functions $b^+$ also close an algebra. The restricted Schur basis corresponds to what goes in the literature under the name of `spherical functions'. These functions are defined as 
\begin{equation}\label{definitionomega}
\omega_R(\sigma)=\frac{1}{|S_n[S_2]|}\sum_{\xi\in S_n[S_2]}\chi_{2R}(\xi\sigma), \quad \sigma\in S_{2n},\quad R\vdash n.
\end{equation} 
We see that $\omega_R(\xi\sigma\eta)=\omega_R(\sigma)$ for all $\eta, \xi\in S_n[S_2]$, as required. We can apply the theorem as a test in this case. Functions $\omega$ are $\Delta$, $H_1=H_2=S_n[S_2]$ and $\psi_1=\psi_2$ are the trivial characters of $S_n[S_2]$. We know that (it is a Littlewood's result) $1\!\!\uparrow_{S_n[S_2]}^{S_{2n}}$ is a multiplicity-free sum of irreps of $S_{2n}$ with even number of boxes in each row. So, the space $\text{Hom}_G(\psi_1^G,\psi_2^G)$ is (by Schur Lemma) the set of maps $2R\to 2R$ for all  $R\vdash n$. These maps can obviously be labeled by $R\vdash n$, and so can the spherical functions since they are in one-to-one correspondence. So the number of spherical functions that form the basis matches the theorem's prediction.\\
Now, from the orthogonality of characters we see that spherical functions have the convolution relations
\begin{equation*}
\omega_R\star\omega_S~(\sigma)=\delta_{RS}\frac{|S_{2n}|}{d_{2R}}\omega_R(\sigma).
\end{equation*}
 Again, we will take the normalization
\begin{equation*}
b^+_R\equiv\frac{d_{2R}}{|S_{2n}|}\omega_R
\end{equation*}
To get the relations
\begin{equation*}
b^+_R\star b^+_S=\delta_{RS}b^+_R,\quad R,S\vdash n,
\end{equation*}
which completely determine functions $b^+_R$.\\
Of course when we consider functions of $\rho\in S_{2\lambda}\subset S_{2n}$, because the spherical function of a product of representations is basically a character of a product of representations we can, as in the unitary case, name $\mu=(r_1\vdash n_1, \dots, r_{l(\lambda)}\vdash n_{l(\lambda)})$ irrep of $S_\lambda$, and $d_\mu=d_{r_1}\cdots d_{l(\lambda)}$, so
\begin{equation*}
\omega_\mu(\rho)=\frac{1}{|S_\lambda[S_2]|}\sum_{\xi\in S_\lambda[S_2]}\chi_{2\mu}(\xi\rho),\quad \rho\in S_{2\lambda}
\end{equation*}
which can be normalized as
\begin{equation*}
b^+_\mu=\frac{d_{2\mu}}{|S_n[S_2]|}\omega_\mu
\end{equation*}
to obtain the relations
\begin{equation*}
b^+_\mu\star b^+_\nu=\delta_{\mu\nu}b^+_\mu.
\end{equation*}

\paragraph{Functions $b^+$ of $S_{2n}$ with $\eta, \xi\in S_\lambda[S_2]$.}
These are functions of $S_{2n}$ that have the symmetry $b^+(\xi\sigma\eta)=b^+(\sigma)$ for all $\eta, \xi\in S_\lambda[S_2]$. This case is similar to the restricted charater case in the unitary group as we are going to see. Let us first discuss the labels of the basis. Referring to the theorem, in this case our linear characters are the trivial ones but of $S_\lambda[S_2]$. So, the labels for our functions $b^+$ are going to be in one-to-one correspondence with   $\text{Hom}_{S_{2n}}(1\!\!\uparrow_{S_\lambda[S_2]}^{S_{2n}},1\!\!\uparrow_{S_\lambda[S_2]}^{S_{2n}})$. But we know what this space is. Because induction is transitive we can perform first the induction $1\!\!\uparrow_{S_\lambda[S_2]}^{S_{2\lambda}}$, where $S_{2\lambda}$ is understood as the group $S_{2n_1}\times S_{2n_2}\times...$, and then induce the resulting representation up to $S_{2n}$. After the first induction we find the direct sum of all irreps  $2\mu=(2r_1\vdash 2n_1, \dots, 2r_{l(\lambda)}\vdash 2n_{l(\lambda)})$ of $S_{2\lambda}$. After the second induction we get a sum of all irreps $R$ of $S_{2n}$ with their multiplicities (if any) that come from the
product $2r_1\times\cdots \times  2r_{l(\lambda)}$. By Frobenius reciprocity we can think of this space as the set of homomorphisms
\begin{equation}\label{bplusmaps}
(R,\mu,ij):~R,\mu,i\to R,\mu,j, \qquad R\vdash 2n,\quad \mu \text{ irrep of } S_\lambda,\quad i,j=1,\dots,g(R;2\mu)
\end{equation}
where $R$ is an irrep of $S_{2n}$,  arbitrary as long as it subduces  $2\mu$ when restricted to $S_{2\lambda}$. Labels $i,j$ run over the multiplicities, the copies of $2\mu$ that come out from the subduction. The number of copies is given by the Littlewood-Richardson coefficient  $g(R;2\mu)$. Note that, by Schur Lemma, irrep $\mu$ must be the same in both sides of the homomorphism, but the multiplicities need not, because one can always establish a non-trivial  homomorphism between two different copies of the same irrep. The dimension of this space of homomorphisms is easily calculated to be
\begin{equation*}
\sum_{R,\mu}g(R;2\mu)^2
\end{equation*}
which must coincide with the dimension of the space of functions  $\{b_{R,\mu,ij}^+\}$.\\
In short, our basis will be labeled as $\{b^+_{R,\mu,ij}\}$, where $R$ is an  irrep of $S_{2n}$, $\mu$ an irrep of $S_\lambda$ and $i,j=1,2,\dots g(R;2\mu)$. \\
Now, spherical functions are basically characters. The sum in the hyperoctahedral group $S_n[S_2]$ that appears in their definition can be interpreted as a projector $P_{[S]}$ (acting on the carrier space of $R$) onto the trivial representation of $S_n[S_2]$ 
\begin{equation*}
\frac{1}{|S_n[S_2]|}\sum_{\xi\in S_n[S_2]}\chi_{2R}(\xi \sigma)=\text{Tr}(P_{[S]}\Gamma_{2R}(\sigma)).
\end{equation*}
To construct the `restricted' spherical functions $\{b^+_{R,\mu,ij}\}$, we can use the same technology as for restricted characters in the unitary case. Irrep $[S]$ will be the trivial representation of $S_\lambda [S_2]$. The fact that $S_\lambda[S_2]\subset S_{2\lambda}$ indicates that the intertwiners $P_{R,\mu,ij}$ as defined in (\ref{propertiesprojectors}), and in particular $P_{R,2\mu,ij}$,  will commute with $P_{[S]}$. So, they will serve to construct  functions 
\begin{equation*}
b^+_{R,\mu,ij}(\sigma)= \frac{d_R}{|S_{2n}|}\text{Tr}(P_{[S]}P_{R,2\mu,ij}\Gamma_R(\sigma)),
\end{equation*}
that fulfill the algebra relations
\begin{equation}\label{restrictedsphericalfunctionsrelations}
b^+_{R,\mu,ij}\star b^+_{S,\nu,kl}=\delta_{RS}\delta_{\mu\nu}\delta_{jk}b^+_{R,\mu,il}.
\end{equation}
As in the case of restricted characters, although  these relations fully characterize the restricted spherical functions, they are not very useful for our applications. A first look at (\ref{restrictedsphericalfunctionsrelations}) reveals 
\begin{equation*}
\delta^+_\lambda=\sum_{R\vdash 2n}\sum_{\mu \text{ irrep of }S_\lambda}\sum_{i=1}^{g(R,2\mu)}b^+_{R,\mu,ii},
\end{equation*}
since it makes $\delta^+_\lambda\star b^+_{R,\mu,ij}=b^+_{R,\mu,ij}\star \delta^+_\lambda= b^+_{R,\mu,ij}$.\\
We can find analogous commutative relations to (\ref{usefulrestricted}) for restricted spherical functions:
\begin{eqnarray}\label{usefulspherical}
b^+_R\star b^+_{S,\mu,ij}&=&\delta_{S\, 2R}b^+_{S,\mu,ij},\nonumber \\
\dot{b}^+_{\mu} \star b^+_{S,\nu,ij}&=&\delta_{\mu\nu} b^+_{S,\nu,ij}, \nonumber \\
b^+_R \star \dot{b}^+_\mu &=&\sum_i  b^+_{2R,\mu,ii},
\end{eqnarray}
with $R\vdash n$ and $S\vdash 2n$.

\paragraph{Functions $b^{SO}$.}
The symmetries of the multitrace monomials in CFT's with orthogonal gauge groups for $\lambda\vdash n$ field content, are
\begin{equation*}
\text{Tr}(\eta\sigma\xi \Psi)=\text{Tr}(\sigma \Psi)\text{sgn}(\eta), \qquad \sigma\in S_{2n},\quad \eta\in S_\lambda[S_2],\quad \xi\in S_n[S_2].
\end{equation*}
So, we are facing to study functions of $\sigma\in S_{2n}$ that fulfill
\begin{equation}\label{bisphericalproperty}
b^{SO}(\eta\sigma\xi)=b^{SO}(\sigma)\text{sgn}(\eta), \quad \eta\in S_\lambda[S_2],\quad \xi\in S_n[S_2].
\end{equation}
For half-BPS operators, where $\lambda=(n)$, these functions where first studied in \cite{I} and baptized as bispherical functions. For this reason we will call the basis of functions which behave as in (\ref{bisphericalproperty}) restricted bispherical functions. How many of these functions will form a basis? How should we label them? To answer these questions we refer again to the theorem (\ref{theorem}). Let as call $[A]$ the sign representation of $S_\lambda[S_2]$ and $[S]$ the trivial representation of $S_n[S_2]$. The space of functions as a vector space will be isomorphic to the space  $\text{Hom}_{S_{2n}}([A]\!\!\uparrow_{S_\lambda[S_2]}^{S_{2n}},[S]\!\!\uparrow_{S_n[S_2]}^{S_{2n}})$. It is a Littlewood result that $[A]\!\!\uparrow_{S_n[S_2]}^{S_{2n}}$ is the multiplicity-free direct sum of irreps of $S_{2n}$ with even number of boxes in each column. So, 
\begin{equation}\label{[A]}
[A]\!\!\uparrow_{S_\lambda[S_2]}^{S_{2\lambda}}=\oplus_\mu (\mu\cup\mu),\quad \mu \text{ irrep of } S_\lambda.
\end{equation}
And, as before
\begin{equation*}
[S]\!\!\uparrow_{S_n[S_2]}^{S_{2n}}=\oplus_R (2R),\quad R\vdash n.
\end{equation*}
Note that in this case, the multiplicities appear only in one side of the homomorphism, namely when performing the second induction
$\uparrow_{S_{2\lambda}}^{S_{2n}}$ in (\ref{[A]}). For this reason the space of homomorphisms (and thus the space of functions)  will have a basis labeled by one multiplicity index:
\begin{equation*}
(R,\mu,i):~\mu\to R,\mu,i, \qquad R\vdash n,\quad \mu \text{ irrep of } S_\lambda,\quad i=1,\dots,g(2R;\mu\cup\mu).
\end{equation*}
So, valid labels for restricted bispherical functions are the ones for which the multiplicity index is not 0, so irrep $2R$ of $S_{2n}$ must subduce at least once the irrep $\mu\cup\mu$ of $S_{2\lambda}$. They can be easily counted as
\begin{equation*}
 \text{Dim } \text{Hom}_{S_{2n}}([A]\!\!\uparrow_{S_\lambda[S_2]}^{S_{2n}},[S]\!\!\uparrow_{S_n[S_2]}^{S_{2n}})=\text{Card}\{b^{SO}_{R,\mu,i}\}=\sum_{R,\mu}g(2R;\mu\cup\mu).
\end{equation*}
This result is in agreement with the counting by means of the evaluation of the partition function for large $N$ in \cite{GK1}.\\
According to the construction we are giving for restricted characters, $\{b^{SO}\}$ will be a set of functions of $S_{2n}$ of the type
\begin{equation}\label{bisphericalschurs}
b^{SO}_{R,\mu,i}(\sigma)=\text{Tr}(P_{2R,\mu\cup\mu,i}\Gamma_{2R}(\sigma)) \quad R\vdash n,\quad \mu \text{ irrep of }S_\lambda,
\end{equation}
where the objects $P_{2R,\mu\cup\mu,i}$ will be intertwiners/projectors acting on the carrier space of $\Gamma_R$. The construction of such objects is out of the scope of this paper. See \cite{GK1} for details. We are going to offer some reasonable relations that, without being too speculative, these functions must fulfill. Since $b^O\star b^O=0~~\forall b^O$, we will partially characterize (up to multiplicities)  $b^{SO}$  by the right convolution product with functions $b^+$. We declare that 
\begin{eqnarray*}
b^{SO}_{R,\mu,i}\star b^+_S&=&\delta_{RS}b^{SO}_{R,\mu,i}, \quad R,S\vdash n, \label{first}\\
b^{SO}_{R,\mu,i}\star \dot{b}^+_\nu&=&\delta_{\mu\nu} b^{SO}_{R,\mu,i},\quad \mu,\nu\text{ irreps of }S_\lambda, \label{second}
\end{eqnarray*}
or equivalently
\begin{equation}\label{third}
b^{SO}_{R,\mu,i}\star\bigg(\sum_j b^+_{S,\nu,jj}\bigg)=\delta_{2R\,S}\delta_{\mu\nu}b^{SO}_{R,\mu,i}, \quad R\vdash n,\quad S\vdash 2n
\end{equation}
Relations (\ref{first}) are straightforwardly fulfilled from (\ref{bisphericalschurs}) and Schur orthogonality of representations. For the relations (\ref{second}) we will first point out that 
\begin{equation*}
 \sum_{\nu\text{ irrep of }S_\lambda} \dot{b}^+_\nu=\delta_\lambda^+ ~~\text{ and }~~b^{SO}_{R,\mu,i}\star\delta_\lambda^+ =b^{SO}_{R,\mu,i}.
\end{equation*}
So, we are sure that
\begin{equation*}
b^{SO}_{R,\mu,i}\star \bigg( \sum_{\nu\text{ irrep of }S_\lambda}\dot{b}^+_\nu\bigg)=b^{SO}_{R,\mu,i}, 
\end{equation*}
for all $R\vdash n$ and $\mu$ irrep of $S_\lambda$, and then (\ref{second}) feels like reasonable projections.\\
Let us comment the last point in more detail. All along this section we are extracting the essence, say, of restricted characters by means of their algebra relations with the convolution product. Because, such bases have not been explicitly constructed in all cases, we have had to derive relations (\ref{first}) and (\ref{second}) as reasonable ``guesses''. Instead, we could have declare (instead of claiming) that the restricted Schur polynomials (which are driven by restricted bases of functions) come as eigenvectors of the charges $\{Q^{\vdash n}, Q^{\vdash \vdash n}, Q^m\}$, which is true for restricted Schur polynomials in the unitary case, that is, for all examples we explicitly know. Now, in this paper we construct  the charges $\{Q^{\vdash n}, Q^{\vdash \vdash n}\}$, and it turns out (see sections \ref{proofsQn} and \ref{proofQnn}) that (\ref{first}) and (\ref{second}) are necessary and sufficient conditions for restricted operators to be their eigenvalues.  In other words, relations (\ref{first}) and (\ref{second}) could be taken as {\it definitions} of  restricted bases of functions for the orthogonal case, up to multiplicities.

\subsection{Symplectic groups}

For CFT's with symplectic gauge groups gauge invariant operators are generated as linear combinations of multitrace monomials as well. Now, since the fields that build the operators are elements of $\mathfrak{sp}(N)$, multitrace monomials can be written as
\begin{equation}\label{multitracesymplecticdefinition}
\text{Tr}_{Sp(N)}(\sigma \Psi)= J_{\sigma(I)}(J\Psi)^I,\quad \sigma \in S_{2n},
\end{equation}
where  
\begin{equation*} 
(J\Psi)^I=(J\phi_1)^{i_1i_2}\cdots (J\phi_1)^{i_{2n_1-1}i_{2n_1}}(J\phi_2)^{i_{2n_1+1}i_{2n_1+2}}\cdots
\end{equation*}
Note that matrices $J\phi_i$ are symmetric, whereas matrices $J$ are antisymmetric. We see from (\ref{multitracesymplecticdefinition}) that multitrace monomials for symplectic gauge groups have the symmetry
\begin{equation*}
\text{Tr}_{Sp(N)}(\eta\sigma\xi \Psi)=\text{Tr}_{Sp(N)}(\sigma \Psi)\text{sgn}(\xi),\quad \eta\in S_\lambda[S_2], \quad \xi\in S_n[S_2].
\end{equation*}
We will call $b^{Sp}$ the functions of $S_{2n}$ that have the same symmetry, that is,
\begin{equation*}
b^{Sp}(\eta\sigma\xi \Psi)=b^{Sp}(\sigma)\text{sgn}(\xi), \quad \eta\in S_\lambda[S_2], \quad \xi\in S_n[S_2].
\end{equation*}
It is easy to see that $b^{Sp}\star b^+=0$ and $b^{Sp}\star b^{Sp}=0$, whereas $b^{Sp}\star b^-$ is, in general, another function $b^{Sp}$. So, functions $b^{Sp}$ will be characterize by their relations with $b^-$ via the right convolution product. \\
For the algebra of $b^-$ we define the unit as
 \begin{equation}\label{unitminus}
\delta_\lambda^-(\sigma)=\frac{1}{|S_\lambda[S_2]|}\sum_{\xi\in S_\lambda[S_2]}\delta(\xi \sigma)\text{sgn}(\xi),\quad \sigma \in S_{2n}.
\end{equation}
The analysis of functions is completely analogous to that of $b^+$, so we are going to point out the differences and state the results. \\
For the half-BPS case, that is functions with the symmetry
\begin{equation*}
b^-(\eta\sigma\xi)=b^-(\sigma)\text{sgn}(\xi),\quad \eta,\xi\in S_n[S_2]
\end{equation*}
the restricted basis has been studied under the name of `twisted spherical functions':
\begin{equation*}
\omega^\varepsilon_R(\sigma)=\frac{1}{|S_n[S_2]|}\sum_{\eta\in S_n[S_2]}\chi_{R\cup R}(\eta\sigma)\text{sgn}(\eta), \quad \sigma\in S_{2n},\quad R\vdash n. 
\end{equation*}
Choosing the normalization
\begin{equation*}
b^-_R\equiv \frac{d_{R\cup R}}{|S_{2n}|}\omega^\varepsilon_R,
\end{equation*}
we have the relations
\begin{equation*}
b^-_R\star b^-_S=\delta_{RS}b^-_R, \quad R\vdash n.
\end{equation*}
Analogous results to  $b^+$ functions, are obtained for $b^-$ when they are functions of $\rho\in S_{2\lambda}$. Then
\begin{equation*}
\omega^\varepsilon_\mu(\rho)=\frac{1}{|S_\lambda[S_2]|}\sum_{\xi\in S_\lambda[S_2]}\chi_{\mu\cup\mu}(\xi\rho)\text{sgn}(\xi),\quad \rho\in S_{2\lambda}
\end{equation*}
which can be normalized as
\begin{equation*}
b^-_\mu=\frac{d_{\mu\cup\mu}}{|S_n[S_2]|}\omega^\varepsilon_\mu
\end{equation*}
to obtain the relations
\begin{equation*}
b^-_\mu\star b^-_\nu=\delta_{\mu\nu}b^-_\mu.
\end{equation*} 
With this functions we define the extensions $\dot{b}^-_\mu$ of $S_{2n}$ as
\begin{equation*}
\dot{b}^-_\mu(\sigma)= \left\{\begin{array}{rl}
b^-_\mu(\sigma) &\text{if } \sigma\in S_{2\lambda} ,\\
0 & \text{otherwise},
\end{array}\right.
\end{equation*}
And we have the identity
\begin{equation*}
\delta_\lambda^-=\sum_{\mu \text{ irrep of }S_\lambda}\dot{b}^-_\mu.
\end{equation*}
For functions 
\begin{equation*}
b^-(\eta\sigma\xi)=b^-(\sigma)\text{sgn}(\xi), \quad \eta,\xi\in S_\lambda[S_2]
\end{equation*}
we find the same labeling as for $b^+$, but the basis of homomorphisms, as read from the theorem, are
\begin{equation}\label{bminusmaps}
(R,\mu,ij):~R,\mu,i\to R,\mu,j, \qquad R\vdash 2n,\quad \mu \text{ irrep of } S_\lambda,\quad i,j=1,\dots,g(R;\mu\cup \mu),
\end{equation}
so $R$ must subduce irreps of $S_{2\lambda}$ with an even number of boxes in each column.\\
Calling $[A]$ the antisymmetric irrep of $S_\lambda[S_2]$ we see that $P_{R,\mu\cup\mu,ij}$ commutes with $P_{[A]}$, and so we will define
\begin{equation*}
b^-_{R,\mu,ij}(\sigma)= \frac{d_R}{|S_{2n}|}\text{Tr}(P_{[A]}P_{R,\mu\cup\mu,ij}\Gamma_R(\sigma)), \quad R\vdash 2n,\quad \mu \text{ irrep of } S_\lambda,
\end{equation*}
to obtain the algebra relations
\begin{equation}\label{restrictedsphericalfunctionsrelations}
b^-_{R,\mu,ij}\star b^-_{S,\nu,kl}=\delta_{RS}\delta_{\mu\nu}\delta_{jk}b^-_{R,\mu,il}.
\end{equation}

\paragraph{Functions $b^{Sp}$.}
For functions
\begin{equation*}
b^{Sp}(\eta\sigma\xi \Psi)=b^{Sp}(\sigma)\text{sgn}(\xi), \quad \eta\in S_\lambda[S_2], \quad \xi\in S_n[S_2].
\end{equation*}
We will find the labeling from a similar analysis we did with $b^{SO}$. We find that a basis of the  space of homomorphisms 
\begin{equation*}
 \text{Hom}_{S_{2n}}([S]\!\!\uparrow_{S_\lambda[S_2]}^{S_{2n}},[A]\!\!\uparrow_{S_n[S_2]}^{S_{2n}})
\end{equation*}
can be labeled as 
\begin{equation*}
(R,\mu,i):~\mu\to R,\mu,i, \qquad R\vdash n,\quad \mu \text{ irrep of } S_\lambda,\quad i=1,\dots,g(R\cup R;2\mu),
\end{equation*}
and we will reasonably define/claim that the algebra relations they satisfy are
\begin{eqnarray*}
b^{Sp}_{R,\mu,i}\star b^-_S&=&\delta_{RS}b^{Sp}_{R,\mu,i}, \quad R,S\vdash n, \label{firstSp}\\
b^{Sp}_{R,\mu,i}\star \dot{b}^-_\nu&=&\delta_{\mu\nu} b^{Sp}_{R,\mu,i},\quad \mu,\nu\text{ irreps of }S_\lambda \label{secondSp}
\end{eqnarray*}
or equivalently
\begin{equation}\label{thirdSp}
b^{Sp}_{R,\mu,i}\star\bigg(\sum_j b^-_{S,\nu,jj}\bigg)=\delta_{R\cup R\,S}\delta_{\mu\nu}b^{Sp}_{R,\mu,i},\quad R\vdash n,\quad S\vdash 2n.
\end{equation}

Relations (\ref{firstSp}) and (\ref{secondSp}) are very similar to (\ref{first}) and (\ref{second}). Indeed $b^O$ and $b^{Sp}$ are identical in the half-BPS case\cite{Pablo}, where the functions are invariant up to a sign in the double coset $S_n[S_2]\backslash S_{2n}/S_n[S_2]$.

\section{General proofs for $Q^{\vdash n}_{NM}$}\label{proofsQn}
Charges $Q^{\vdash n}_{NM}$ acts naturally on gauge invariant operators built on $\Psi$ in much the same way as it acts on the half-BPS sector where only one scalar matrix is considered. This operator is self-adjoint by construction but we can check it.
\subsection{Self-adjointness}

 First, we find that when they act on multitrace monomials they give
\begin{eqnarray}\label{Qnontraces} 
Q^{\vdash n}_{NM}\big[\text{Tr}_{U(N)}(\sigma\Psi)\big]&=&\frac{1}{|S_n|}\sum_{\beta\in S_n}\sum_{\substack{R\vdash n\\
l(R)\leq N}}d_R\frac{f_R^{U(N)}}{f_R^{U(M)}}\chi_R(\beta^{-1}\sigma)\text{Tr}_{U(N)}(\beta\Psi)\nonumber \\
Q^{\vdash n}_{NM}\big[\text{Tr}_{SO(N)}(\sigma\Psi)\big]&=&\frac{1}{|S_{2n}|}\sum_{\beta\in S_{2n}}\sum_{\substack{R\vdash n\\
l(R)\leq N}}d_{2R}\frac{f_R^{SO(N)}}{f_R^{SO(M)}}\omega_R(\beta^{-1}\sigma)\text{Tr}_{SO(N)}(\beta\Psi)\nonumber \\
Q^{\vdash n}_{NM}\big[\text{Tr}_{Sp(N)}(\sigma\Psi)\big]&=&\frac{1}{|S_{2n}|}\sum_{\beta\in S_{2n}}\sum_{\substack{R\vdash n\\
l(R)\leq N}}d_{R\cup R}\frac{f_R^{Sp(N)}}{f_R^{Sp(M)}}\omega^\varepsilon_R(\beta^{-1}\sigma)\text{Tr}_{Sp(N)}(\beta\Psi),
\end{eqnarray}
where $\sigma\in S_n$ for unitary groups and $\sigma\in S_{2n}$ for orthogonal and symplectic groups. For convenience let us write
(\ref{Qnontraces}) in terms of the functions $b,b^+$ and $b^-$ defined in section  (\ref{CRBCP}). We have
\begin{eqnarray}\label{Qnontraceswithb} 
Q^{\vdash n}_{NM}\big[\text{Tr}_{U(N)}(\sigma\Psi)\big]&=&\sum_{\beta\in S_n}\sum_{\substack{R\vdash n\\
l(R)\leq N}}\frac{f_R^{U(N)}}{f_R^{U(M)}}b^U_R(\beta^{-1}\sigma)\text{Tr}_{U(N)}(\beta\Psi)\nonumber \\
Q^{\vdash n}_{NM}\big[\text{Tr}_{SO(N)}(\sigma\Psi)\big]&=&\sum_{\beta\in S_{2n}}\sum_{\substack{R\vdash n\\
l(R)\leq N}}\frac{f_R^{SO(N)}}{f_R^{SO(M)}}b^+_R(\beta^{-1}\sigma)\text{Tr}_{SO(N)}(\beta\Psi)\nonumber \\
Q^{\vdash n}_{NM}\big[\text{Tr}_{Sp(N)}(\sigma\Psi)\big]&=&\sum_{\beta\in S_{2n}}\sum_{\substack{R\vdash n\\
l(R)\leq N}}\frac{f_R^{Sp(N)}}{f_R^{Sp(M)}}b^-_R(\beta^{-1}\sigma)\text{Tr}_{Sp(N)}(\beta\Psi).
\end{eqnarray}
In order to prove self-adjointness of $Q^{\vdash n}_{NM}$ we must see that 
\begin{equation*}
\langle Q^{\vdash n}_{NM}\big[\text{Tr}_{G(N)}(\sigma\Psi)\big]\text{Tr}_{G(N)}(\tau\bar{\Psi})\rangle
\end{equation*}
is invariant under the exchange $\sigma\leftrightarrow \tau$.\\
 We will need to know how Wick contractions go for multitrace monomials. This is written in equations (\ref{correlatorsallgauge}). We reproduce it here in terms of normalized functions $b$. For the unitary group we have
\begin{equation}\label{wicksmultitracesunitary}
\langle\text{Tr}_{U(N)}(\beta \Psi) \text{Tr}_{U(N)}(\tau \bar{\Psi})\rangle
=\sum_{R\vdash n}\sum_{\rho\in S_\lambda} f^{U(N)}_R b^U_R(\beta^{-1}\rho\tau\rho^{-1}),
\end{equation}
for orthogonal groups
\begin{equation}\label{wicksmultitraceorthogonal}
\langle\text{Tr}_{SO(N)}(\beta \Psi) \text{Tr}_{SO(N)}(\tau \bar{\Psi})\rangle
=|S_n[S_2]|\sum_{R \vdash n}\sum_{\eta\in S_\lambda[S_2]}f^{SO(N)}_{R}b^+_{R}(\beta^{-1}\eta\tau)\text{sgn}(\eta)
\end{equation}
and for symplectic groups
\begin{equation}\label{wicksmultitracesymplectic}
\langle\text{Tr}_{Sp(N)}(\beta \Psi) \text{Tr}_{Sp(N)}(\tau \bar{\Psi})\rangle
=|S_n[S_2]|\sum_{R \vdash n}\sum_{\eta\in S_\lambda[S_2]}f^{Sp(N)}_{R}b^-_{R}(\beta^{-1}\eta\tau)\text{sgn}(\eta).
\end{equation}
The key point here is that functions $b_R(\sigma)=b_R(\sigma^{-1})$, because they are essentially characters. The same happens with $b^+_R$ and $b^-_R$ because they are invariant over the elements of the double coset $S_n[S_2]\backslash S_{2n}/S_n[S_2]$, and $\sigma^{-1}\in S_{2n}$ belongs to the double coset of $\sigma\in S_{2n}$.\\
Now, for the unitary groups we have
\begin{eqnarray*}
&&\langle Q^{\vdash n}_{NM}\big[\text{Tr}_{U(N)}(\sigma\Psi)\big]\text{Tr}_{U(N)}(\tau\bar{\Psi})\rangle \nonumber \\
&=&\sum_{\beta\in S_n}\sum_{\substack{R\vdash n\\
l(R)\leq N}}\frac{f_R^{U(N)}}{f_R^{U(M)}}b^U_R(\beta^{-1}\sigma)\langle \text{Tr}_{U(N)}(\beta\Psi)\text{Tr}_{U(N)}(\beta\bar{\Psi})\rangle \nonumber\\
&=&\sum_{\beta\in S_n}\sum_{\substack{R\vdash n\\
l(R)\leq N}}\sum_{S\vdash n}\sum_{\rho\in S_\lambda}\frac{f_R^{U(N)}}{f_R^{U(M)}}f^{U(N)}_S b^U_R(\beta^{-1}\sigma)  b^U_S(\beta^{-1}\rho\tau\rho^{-1}) \nonumber \\
&=&\sum_{\beta\in S_n}\sum_{\substack{R\vdash n\\
l(R)\leq N}}\sum_{S\vdash n}\sum_{\rho\in S_\lambda}\frac{f_R^{U(N)}}{f_R^{U(M)}}f^{U(N)}_S b^U_R(\beta^{-1})  b^U_S(\beta^{-1}\sigma^{-1}\rho\tau\rho^{-1}) \nonumber \\
&=&\sum_{\substack{R\vdash n\\
l(R)\leq N}}\sum_{\rho\in S_\lambda}\frac{\big(f_R^{U(N)}\big)^2}{f_R^{U(M)}}  b^U_R(\sigma^{-1}\rho\tau\rho^{-1}),
\end{eqnarray*}
which is clearly invariant under the swap $\sigma\leftrightarrow \tau$ because $b^U_R(\sigma)=b^U_R(\sigma^{-1})$. \\
For the orthogonal case we obtain
\begin{equation*}
\langle Q^{\vdash n}_{NM}\big[\text{Tr}_{SO(N)}(\sigma\Psi)\big]\text{Tr}_{SO(N)}(\tau\bar{\Psi})\rangle=|S_n[S_2]|\sum_{\substack{R\vdash n\\
l(R)\leq N}}\sum_{\eta\in S_\lambda[S_2]}\frac{\big(f_R^{SO(N)}\big)^2}{f_R^{SO(M)}}  b^+_R(\sigma^{-1}\eta\tau)\text{sgn}(\eta),
\end{equation*}
which is again invariant under the swap $\sigma\leftrightarrow \tau$ because $b^+_R(\sigma)=b^+_R(\sigma^{-1})$.\\
Similar result is found for the symplectic case:
\begin{equation*}
\langle Q^{\vdash n}_{NM}\big[\text{Tr}_{Sp(N)}(\sigma\Psi)\big]\text{Tr}_{Sp(N)}(\tau\bar{\Psi})\rangle=|S_n[S_2]|\sum_{\substack{R\vdash n\\
l(R)\leq N}}\sum_{\eta\in S_\lambda[S_2]}\frac{\big(f_R^{Sp(N)}\big)^2}{f_R^{Sp(M)}}  b^-_R(\sigma^{-1}\eta\tau)\text{sgn}(\eta),
\end{equation*}
which also invariant  under the swap $\sigma\leftrightarrow \tau$ since $b^-_R(\sigma)=b^-_R(\sigma^{-1})$.\\

\subsection{Eigenvectors}

To see that restricted Schur polynomials are eigenvectors of $Q^{\vdash n}_{NM}$ we must remember the algebra of functions $b,b^+$ and $b^-$ as shown in section (\ref{CRBCP}). Specifically
\begin{eqnarray*}
b^{U}_{S,\mu,ij}\star b^U_R&=&\delta_{RS}b^{U}_{S,\mu,ij},\nonumber \\
b^{SO}_{S,\mu,i} \star b^+_R &=&\delta_{RS}b^{SO}_{S,\mu,i},\nonumber \\
b^{Sp}_{S,\mu,i}\star  b^-_R&=&\delta_{RS}b^{Sp}_{S,\mu,i},\nonumber
\end{eqnarray*}
 Remember that our restricted Schur polynomials are defined as
\begin{equation*}
\chi^{G(N)}_{R,\mu,m}(\Psi)=\sum_{\sigma\in S_n(S_{2n})}b^{G}_{R,\mu,m}(\sigma)\text{Tr}_{G(N)}(\sigma\Psi).
\end{equation*} 
Now, using (\ref{Qnontraceswithb}) we have for the unitary case
\begin{eqnarray*}
Q^{\vdash n}_{NM}\big[\chi^{U(N)}_{R,\mu,ij}(\Psi)\big]&=&\sum_{\beta\in S_n}\sum_{\substack{S\vdash n\\
l(S)\leq N}}\frac{f_S^{U(N)}}{f_S^{U(M)}}b^U_S(\beta^{-1}\sigma)b^{U}_{R,\mu,ij}(\sigma)\text{Tr}_{U(N)}(\beta\Psi)\nonumber \\
&=&\sum_{\beta\in S_n}\sum_{\substack{S\vdash n\\
l(S)\leq N}}\frac{f_S^{U(N)}}{f_S^{U(M)}}b^U_S(\sigma^{-1}\beta)b^{U}_{R,\mu,ij}(\sigma)\text{Tr}_{U(N)}(\beta\Psi)\nonumber \\
&=&\sum_{\beta\in S_n}\frac{f_R^{U(N)}}{f_R^{U(M)}}b^{U}_{R,\mu,ij}(\beta)\text{Tr}_{U(N)}(\beta\Psi)\nonumber \\
&=& \frac{f_R^{U(N)}}{f_R^{U(M)}}\chi^{U(N)}_{R,\mu,ij}(\Psi).
\end{eqnarray*}
For orthogonal gauge groups we have
\begin{eqnarray*}
Q^{\vdash n}_{NM}\big[\chi^{SO(N)}_{R,\mu,i}(\Psi)\big]&=&\sum_{\beta\in S_{2n}}\sum_{\substack{S\vdash n\\
l(S)\leq N}}\frac{f_S^{SO(N)}}{f_S^{SO(M)}}b^+_S(\beta^{-1}\sigma)b^{SO}_{R,\mu,i}(\sigma)\text{Tr}_{SO(N)}(\beta\Psi)\nonumber \\
&=& \frac{f_R^{SO(N)}}{f_R^{SO(M)}}\chi^{SO(N)}_{R,\mu,i}(\Psi).
\end{eqnarray*}
And for symplectic gauge groups
\begin{eqnarray*}
Q^{\vdash n}_{NM}\big[\chi^{Sp(N)}_{R,\mu,i}(\Psi)\big]&=&\sum_{\beta\in S_{2n}}\sum_{\substack{S\vdash n\\
l(S)\leq N}}\frac{f_S^{Sp(N)}}{f_S^{Sp(M)}}b^+_S(\beta^{-1}\sigma)b^{Sp}_{R,\mu,i}(\sigma)\text{Tr}_{Sp(N)}(\beta\Psi)\nonumber \\
&=& \frac{f_R^{Sp(N)}}{f_R^{Sp(M)}}\chi^{Sp(N)}_{R,\mu,i}(\Psi).
\end{eqnarray*}

\section{General proof for $Q^{\vdash\vdash n}_{NM}$}\label{proofQnn}
We have defined
\begin{equation*}
Q^{\vdash\vdash n}_{NM}\equiv\sum_{\substack{\lambda\vdash n\\
\lambda\neq (n)}}Q^\lambda_{NM}=\sum_{\substack{\lambda\vdash n\\
\lambda\neq (n)}}\text{Proj}_{NM}\circ \text{Av}_{MN}^{\lambda}.
\end{equation*}
We will use this definition to prove the properties of $Q^{\vdash\vdash n}_{NM}$ described in section \ref{Qnn}. We will first find how  $Q^{\lambda}_{NM}$ act on multitrace monomials. Using this result and the properties of restricted Schur characters as described in section \ref{CRBCP}, we will prove the self-adjointness of $Q^{\vdash\vdash n}_{NM}$ and find that their eigenvectors are precisely restricted Schur polynomials.

\subsection{$\lambda$-averaging acting on multitrace monomials}

The $\lambda$-averaging operator with the adjoint actions defined in section \ref{Qnn}  have the form
\begin{equation}\label{avinunitary}
\text{Av}_{MN}^{\lambda}[\mathcal{O}_{G(N)}(\Psi)]\equiv\int_{g_1,\dots,g_r\in G(M)}\text{d}\big[g\big]\text{Ad}^{\lambda}_{MN}[\mathcal{O}_{G(N)}(\Psi)].
\end{equation}
We are going to see how it acts on multitrace monomials for each gauge group. 

\paragraph{Unitary gauge groups.}
In order to see how $\text{Av}_{MN}^{\lambda}$ acts on multitrace monomials we should remember the result of the integration of group entries in the unitary case\cite{C}
\begin{equation*}
\int_{g\in U(M)}\text{d}g~g^{i_1}_{j_1}\cdots g^{i_n}_{j_n} (\bar{g})^{i'_1}_{j'_1}\cdots (\bar{g})^{i'_n}_{j'_n}=\sum_{\alpha,\beta\in S_n}(\alpha)^I_{I'}(\beta)^{J'}_J\text{Wg}^{U(M)}(\alpha\beta),
\end{equation*} 
where 
\begin{equation*}
\text{Wg}^{U(M)}(\sigma)=\frac{1}{|S_n|}\sum_{\substack{R\vdash n \\
l(R)\leq M}} \frac{d_R}{f_R^{U(M)}}\chi_R(\sigma) ,\quad \sigma\in S_n.
\end{equation*}
Note that in (\ref{avinunitary}) there are $r$ integrals over the subgroups $S_{n_1},S_{n_2},\dots, S_{n_r}$. We will call
\begin{equation*}
\text{Wg}_\lambda^{U(M)}(\rho)=\frac{1}{|S_\lambda|}\sum_{\substack{\mu\text{ irrep of }S_\lambda \\
l(\mu)\leq M}} \frac{d_\mu}{f_\mu^{U(M)}}\chi_\mu(\rho) ,\quad \rho\in S_\lambda,
\end{equation*}
where we have $\mu=(s_1,\dots, s_r)$ for $s_i\vdash n_i$, and we have defined $d_\mu=d_{s_1}d_{s_2}\cdots d_{s_r}$, $f_\mu^{U(M)}=f_{s_1}^{U(M)}f_{s_2}^{U(M)}\cdots f_{s_r}^{U(M)}$ and $\chi_\mu(\rho)=\chi_{s_1\times\cdots\times s_r}(\rho)$.\\
With this notation we will write
\begin{eqnarray*}
\text{Av}_{MN}^{\lambda}[\text{Tr}_{U(N)}(\sigma\Psi)]&=&\int_{g_1,\dots,g_r\in U(M)}\text{d}\big[g\big]\text{Ad}^{\lambda}_{MN}[\text{Tr}_{U(N)}(\sigma\Psi)] \nonumber \\
&=& \frac{1}{|S_n|}\sum_{\alpha \in S_n} \int_{g_1,\dots,g_r\in U(M)}\text{d}\big[g\big]\big[g\big]^I_{\alpha(J)} \big[\Psi\big]^J_{J'} \big[\bar{g}\big]^{I'}_{J'}(\alpha^{-1}\sigma)^{I'}_I \nonumber \\
&=& \frac{1}{|S_n|}\sum_{\alpha \in S_n} \sum_{\rho_1,\rho_2 \in S_\lambda}(\rho_1)^I_{I'}(\alpha^{-1}\sigma)^{I'}_I(\alpha\rho_2)^{J'}_J \big[\Psi\big]^J_{J'} \text{Wg}_\lambda^{U(M)}(\rho_1\rho_2)\nonumber \\
&=& \frac{1}{|S_n|}\sum_{\alpha \in S_n} \sum_{\rho_1,\rho_2 \in S_\lambda}\text{Tr}_{U(N)}(\rho_1\alpha^{-1}\sigma)\text{Wg}_\lambda^{U(M)}(\rho_1\rho_2)\text{Tr}_{U(M)}(\alpha\rho_2\Psi)  \nonumber \\
&=& \frac{|S_\lambda|}{|S_n|}\sum_{\alpha \in S_n} \sum_{\rho \in S_\lambda}\text{Tr}_{U(N)}(\rho\alpha^{-1}\sigma)\text{Wg}_\lambda^{U(M)}(\rho)\text{Tr}_{U(M)}(\alpha\Psi), 
\end{eqnarray*}
where we have used simple algebra of indeces, see appendix \ref{awi}. Now, for the unitary case we have
\begin{equation*}
Q^\lambda_{MN}\text{Tr}_{U(N)}(\sigma\Psi)=\frac{|S_\lambda|}{|S_n|}\sum_{\alpha \in S_n} \sum_{\rho \in S_\lambda}\text{Tr}_{U(N)}(\rho\alpha^{-1}\sigma)\text{Wg}_\lambda^{U(M)}(\rho)\text{Tr}_{U(N)}(\alpha\Psi).
\end{equation*}

\paragraph{Orthogonal gauge groups}
The orthogonal case presents some minor variations. The integral of the entries of the orthogonal group is\cite{CM}
\begin{equation*}
\int_{g\in O(M)}\text{d}g~g_{i_1j_1}\cdots g_{i_{2n}j_{2n}}=\frac{1}{|S_n[S_2]|^2}\sum_{\alpha,\beta\in S_{2n}}\delta_{\alpha(I)}\delta_{\beta(J)}\text{Wg}^{O(M)}(\alpha^{-1}\beta),
\end{equation*}
with
\begin{equation*}
\text{Wg}^{O(M)}(\sigma)=\frac{|S_n[S_2]|}{|S_{2n}|}\sum_{\substack{R\vdash n \\
l(R)\leq M}} \frac{d_{2R}}{f_R^{O(M)}}\omega_R(\sigma) ,\quad \sigma\in S_{2n},
\end{equation*}
where we have defined 
\begin{equation}\label{fRO}
f_R^{SO(M)}=\prod_{(i,j)\in R}(M+2j-i-1)=Z_{R}({\bf 1}_M).
\end{equation}
Again, we have to perform $r$ integrals in (\ref{avinunitary}) over the subgroups $S_{2n_1},S_{2n_2},\dots,S_{2n_r}$. Using analogous notation as for the unitary case, we will call $\mu=(s_1,\dots, s_r)$ an irrep of $S_\lambda$, $2\mu=(2s_1,\dots, 2s_r)$ an irrep of $S_{2\lambda}$ and define
\begin{equation*}
\text{Wg}_\lambda^{O(M)}(\rho)=\frac{|S_\lambda[S_2]|}{|S_{2\lambda}|}\sum_{\substack{\mu\text{ irrep of }S_\lambda \\
l(\mu)\leq M}} \frac{d_{2\mu}}{f_\mu^{O(M)}}\omega_\mu(\rho) ,\quad \rho\in S_{2\lambda}.
\end{equation*} 
With this notation we will write
\begin{eqnarray*}
\text{Av}_{MN}^{\lambda}[\text{Tr}_{SO(N)}(\sigma\Psi)]&=&\int_{g_1,\dots,g_r\in SO(M)}\text{d}\big[g\big]\text{Ad}^{\lambda}_{MN}[\text{Tr}_{SO(N)}(\sigma\Psi)] \nonumber \\
&=& \frac{1}{|S_{2n}|}\sum_{\alpha \in S_{2n}} \int_{g_1,\dots,g_r\in SO(M)}\text{d}\big[g\big]\big[g\big]_{I\alpha(J)}\Psi^J \delta_{\alpha^{-1}\sigma(I)} \nonumber \\
&=& \frac{1}{|S_{2n}||S_\lambda[S_2]|^2}\sum_{\alpha \in S_{2n}} \sum_{\rho_1,\rho_2 \in S_{2\lambda}}\delta_{\rho_1(I)}\delta_{\alpha^{-1}\sigma(I)}\delta_{\alpha\rho_2(J)}\big[\Psi\big]^J \text{Wg}_\lambda^{SO(M)}(\rho_1^{-1}\rho_2)\nonumber \\
&=& \frac{1}{|S_{2n}||S_\lambda[S_2]|^2}\sum_{\alpha \in S_{2n}} \sum_{\rho_1,\rho_2 \in S_{2\lambda}}\text{Tr}_{SO(N)}(\rho^{-1}_1\alpha^{-1}\sigma)\text{Wg}_\lambda^{SO(M)}(\rho^{-1}_1\rho_2)\text{Tr}_{SO(M)}(\alpha\rho_2\Psi)  \nonumber \\
&=& \frac{|S_{2\lambda}|}{|S_{2n}||S_\lambda[S_2]|^2}\sum_{\alpha \in S_{2n}} \sum_{\rho \in S_{2\lambda}}\text{Tr}_{SO(N)}(\rho\alpha^{-1}\sigma)\text{Wg}_\lambda^{SO(M)}(\rho)\text{Tr}_{SO(M)}(\alpha\Psi), 
\end{eqnarray*}
and so
\begin{equation*}
Q^\lambda_{NM}[\text{Tr}_{SO(N)}(\sigma\Psi)]=\frac{|S_{2\lambda}|}{|S_{2n}||S_\lambda[S_2]|^2}\sum_{\alpha \in S_{2n}} \sum_{\rho \in S_{2\lambda}}\text{Tr}_{SO(N)}(\rho\alpha^{-1}\sigma)\text{Wg}_\lambda^{SO(M)}(\rho)\text{Tr}_{SO(N)}(\alpha\Psi)
\end{equation*}

\paragraph{Symplectic gauge groups}
For symplectic groups the integral on the entries reads\cite{Mat}
\begin{equation*}
\int_{g\in Sp(M)}\text{d}g~g_{i_1j_1}\cdots g_{i_{2n}j_{2n}}=\frac{1}{|S_n[S_2]|^2}\sum_{\alpha,\beta\in S_{2n}}J_{\alpha(I)}J_{\beta(J)}\text{Wg}^{Sp(M)}(\alpha^{-1}\beta),
\end{equation*}
with
\begin{equation*}
\text{Wg}^{Sp(M)}(\sigma)=\frac{|S_n[S_2]|}{|S_{2n}|}\sum_{\substack{R\vdash n \\
l(R)\leq M}} \frac{d_{R\cup R}}{f_R^{Sp(M)}}\omega^{\varepsilon}_R(\sigma) ,\quad \sigma\in S_{2n},
\end{equation*}
where we have defined 
\begin{equation}\label{fRO}
f_R^{Sp(M)}=\prod_{(i,j)\in R}(M+j-2i+1)=Z'_{R}({\bf 1}_{M/2}).
\end{equation}
As we have to perform $r$ integrals  in (\ref{avinunitary}) over the subgroups $S_{2n_1},S_{2n_2},\dots,S_{2n_r}$,  will call $\mu=(s_1,\dots, s_r)$ an irrep of $S_\lambda$, $\mu\cup\mu=(s_1\cup s_1,\dots, s_r\cup s_r)$ an irrep of $S_{2\lambda}$ and define
\begin{equation*}
\text{Wg}_\lambda^{Sp(M)}(\rho)=\frac{|S_\lambda[S_2]|}{|S_{2\lambda}|}\sum_{\substack{\mu\text{ irrep of }S_\lambda \\
l(\mu)\leq M}} \frac{d_{\mu\cup \mu}}{f_\mu^{Sp(M)}}\omega^\varepsilon_\mu(\rho) ,\quad \rho\in S_{2\lambda}.
\end{equation*} 
With this notation we will write
\begin{eqnarray*}
\text{Av}_{MN}^{\lambda}[\text{Tr}_{Sp(N)}(\sigma\Psi)]&=&\int_{g_1,\dots,g_r\in Sp(M)}\text{d}\big[g\big]\text{Ad}^{\lambda}_{MN}[\text{Tr}_{Sp(N)}(\sigma\Psi)] \nonumber \\
&=& \frac{1}{|S_{2n}|}\sum_{\alpha \in S_{2n}} \int_{g_1,\dots,g_r\in Sp(M)}\text{d}\big[g\big]\big[g\big]_{I\alpha(K)}(J\Psi)^K J_{\alpha^{-1}\sigma(I)} \nonumber \\
&=& \frac{1}{|S_{2n}||S_\lambda[S_2]|^2}\sum_{\alpha \in S_{2n}} \sum_{\rho_1,\rho_2 \in S_{2\lambda}}J_{\rho_1(I)}J_{\alpha^{-1}\sigma(I)}J_{\alpha\rho_2(K)}\big[J\Psi\big]^K \text{Wg}_\lambda^{Sp(M)}(\rho_1^{-1}\rho_2)\nonumber \\
&=& \frac{1}{|S_{2n}||S_\lambda[S_2]|^2}\sum_{\alpha \in S_{2n}} \sum_{\rho_1,\rho_2 \in S_{2\lambda}}\text{Tr}_{Sp(N)}(\rho^{-1}_1\alpha^{-1}\sigma)\text{Wg}_\lambda^{Sp(M)}(\rho^{-1}_1\rho_2)\text{Tr}_{Sp(M)}(\alpha\rho_2\Psi)  \nonumber \\
&=& \frac{|S_{2\lambda}|}{|S_{2n}||S_\lambda[S_2]|^2}\sum_{\alpha \in S_{2n}} \sum_{\rho \in S_{2\lambda}}\text{Tr}_{Sp(N)}(\rho\alpha^{-1}\sigma)\text{Wg}_\lambda^{Sp(M)}(\rho)\text{Tr}_{Sp(M)}(\alpha\Psi), 
\end{eqnarray*}
and so
\begin{equation*}
Q^\lambda_{NM}[\text{Tr}_{Sp(N)}(\sigma\Psi)]=\frac{|S_{2\lambda}|}{|S_{2n}||S_\lambda[S_2]|^2}\sum_{\alpha \in S_{2n}} \sum_{\rho \in S_{2\lambda}}\text{Tr}_{Sp(N)}(\rho\alpha^{-1}\sigma)\text{Wg}_\lambda^{Sp(M)}(\rho)\text{Tr}_{Sp(N)}(\alpha\Psi)
\end{equation*}

\subsection{Self-adjointness}

To see that $Q^{\vdash\vdash n}_{NM}$ is self-adjoint it is enough to prove that  $Q^\lambda_{NM}$ is self-adjoint for all $\lambda$. So, what we have to prove is that 
\begin{equation*}
\langle (\text{Proj}_{NM}\circ \text{Av}_{MN}^{\lambda})\text{Tr}_N(\sigma \Psi)\text{Tr}_N(\tau \bar{\Psi})\rangle=\langle \text{Tr}_N(\sigma \Psi) (\text{Proj}_{NM}\circ \text{Av}_{MN}^{\lambda})\text{Tr}_N(\tau \bar{\Psi})\rangle,
\end{equation*}
for all $\sigma,\tau\in S_{n}$ in the unitary case, and $\sigma,\tau\in S_{2n}$ for orthogonal and symplectic gauge groups. For simplicity we will consider the case of operators built on bosonic fields. We will need to know how Wick contractions go for multitrace monomials. Remember that
for  unitary groups we have
\begin{equation}\label{wicksmultitracesunitary}
\langle\text{Tr}_{U(N)}(\sigma \Psi) \text{Tr}_{U(N)}(\tau \bar{\Psi})\rangle
=\sum_{R\vdash n}\sum_{\rho\in S_\lambda} f^{U(N)}_R b^U_R(\sigma^{-1}\rho\tau\rho^{-1}),
\end{equation}
for orthogonal groups
\begin{equation}\label{wicksmultitraceorthogonal}
\langle\text{Tr}_{SO(N)}(\sigma \Psi) \text{Tr}_{SO(N)}(\tau \bar{\Psi})\rangle
=|S_n[S_2]|\sum_{R \vdash n}\sum_{\eta\in S_\lambda[S_2]}Z_{R}({\bf 1}_N)b^+_{R}(\sigma^{-1}\eta\tau)\text{sgn}(\eta)
\end{equation}
and for symplectic groups
\begin{equation}\label{wicksmultitracesymplectic}
\langle\text{Tr}_{Sp(N)}(\sigma \Psi) \text{Tr}_{Sp(N)}(\tau \bar{\Psi})\rangle
=|S_n[S_2]|\sum_{R \vdash n}\sum_{\eta\in S_\lambda[S_2]}Z'_{R}({\bf 1}_{N/2})b^-_{R}(\sigma^{-1}\eta\tau)\text{sgn}(\eta).
\end{equation}
Applying $Q^\lambda_{NM}=\text{Proj}_{NM}\circ \text{Av}_{MN}^{\lambda}$ to multitrace monomials we obtain
\begin{eqnarray*}
Q^\lambda_{NM}\big(\text{Tr}_{U(N)}(\sigma\Psi)\big)&=&\frac{|S_\lambda|}{|S_n|}\sum_{\alpha\in S_n}\sum_{\rho\in S_\lambda}\text{Tr}_{U(N)}(\rho^{-1}\alpha^{-1}\sigma)\text{Wg}_\lambda^{U(M)}(\rho)\text{Tr}_{U(N)}(\alpha\Psi)\nonumber \\
Q^\lambda_{NM}\big(\text{Tr}_{SO(N)}(\sigma\Psi)\big)&=&\frac{|S_\lambda|}{|S_\lambda[S_2]|^2|S_n|}\sum_{\alpha\in S_{2n}}\sum_{\rho\in S_\lambda[S_2]}\text{Tr}_{SO(N)}(\rho^{-1}\alpha^{-1}\sigma)\text{Wg}_\lambda^{SO(M)}(\rho)\text{Tr}_{SO(N)}(\alpha\Psi)\nonumber \\
Q^\lambda_{NM}\big(\text{Tr}_{Sp(N)}(\sigma\Psi)\big)&=&\frac{|S_\lambda|}{|S_\lambda[S_2]|^2|S_n|}\sum_{\alpha\in S_{2n}}\sum_{\rho\in S_\lambda[S_2]}\text{Tr}_{Sp(N)}(\rho^{-1}\alpha^{-1}\sigma)\text{Wg}_\lambda^{Sp(M)}(\rho)\text{Tr}_{Sp(N)}(\alpha\Psi).\nonumber \\
\end{eqnarray*}
Let us compute
\begin{equation*}
\sum_{\rho\in S_\lambda}\text{Tr}_{G(N)}(\rho^{-1}\alpha^{-1}\sigma)\text{Wg}_\lambda^{G(M)}(\rho),
\end{equation*}
for $G=U,SO, Sp$ gauge groups. In order to make the computations agile and clearer, we will make use of convolution properties studied in section (\ref{CRBCP}). For unitary groups we know that 
\begin{equation*}
\text{Tr}_{U(N)}(\rho^{-1}\alpha^{-1}\sigma)=\sum_{R\vdash n}f_R^{U(N)}b^U_R(\rho^{-1}\alpha^{-1}\sigma),
\end{equation*}
and that the Weingarten function of the subgroup $S_\lambda$ can be expanded as
\begin{equation*}
\text{Wg}_\lambda^{U(M)}(\rho)=\frac{1}{|S_\lambda|}\sum_{\mu\text{ irrep of } S_\lambda}d_\mu \frac{1}{f_\mu^{U(M)}}\chi_\mu(\rho)=\sum_{\mu\text{ irrep of } S_\lambda} \frac{1}{f_\mu^{U(M)}}b^U_\mu(\rho).
\end{equation*}
Now, we extend the function to $\beta\in S_{2n}$ by
\begin{equation*}
\text{Wg}_\lambda^{U(M)}(\beta)=\sum_{\mu\text{ irrep of } S_\lambda} \frac{1}{f_\mu^{U(M)}}\dot{b}^U_\mu(\beta),
\end{equation*}
and so
\begin{eqnarray*}
\sum_{\rho\in S_\lambda}\text{Tr}_{U(N)}(\rho^{-1}\alpha^{-1}\sigma)\text{Wg}_\lambda^{U(M)}(\rho)&=&\sum_{\substack{R\vdash n\\
\mu\text{ irrep of } S_\lambda \\
\beta\in S_{2n}}} \frac{f_R^{U(N)}}{f_\mu^{U(M)}}\dot{b}^U_\mu(\beta)b^U_R(\beta^{-1}\alpha^{-1}\sigma)\nonumber \\
&=&\sum_{\substack{R\vdash n\\
\mu\text{ irrep of } S_\lambda}} \frac{f_R^{U(N)}}{f_\mu^{U(M)}}\dot{b}^U_\mu\star b^U_R(\alpha^{-1}\sigma)\nonumber \\
&=&\sum_{\substack{R\vdash n\\
\mu\text{ irrep of } S_\lambda \\
i=1,\dots, g(R;\mu)}} \frac{f_R^{U(N)}}{f_\mu^{U(M)}}b^U_{R,\mu,ii}(\alpha^{-1}\sigma)
\end{eqnarray*}
Performing analogous calculations we find for the orthogonal case
\begin{equation*}
\sum_{\rho\in S_{2\lambda}}\text{Tr}_{SO(N)}(\rho^{-1}\alpha^{-1}\sigma)\text{Wg}_\lambda^{SO(M)}(\rho)=\sum_{\substack{R\vdash n\\
\mu\text{ irrep of } S_\lambda \\
i=1,\dots, g(R;2\mu)}} \frac{f_R^{SO(N)}}{f_\mu^{SO(M)}}b^+_{R,\mu,ii}(\alpha^{-1}\sigma)
\end{equation*}
and for symplectic gauge groups
\begin{equation*}
\sum_{\rho\in S_{2\lambda}}\text{Tr}_{Sp(N)}(\rho^{-1}\alpha^{-1}\sigma)\text{Wg}_\lambda^{Sp(M)}(\rho)=\sum_{\substack{R\vdash n\\
\mu\text{ irrep of } S_\lambda \\
i=1,\dots, g(R;\mu\cup\mu)}} \frac{f_R^{Sp(N)}}{f_\mu^{Sp(M)}}b^-_{R,\mu,ii}(\alpha^{-1}\sigma).
\end{equation*}
So, we will write
\begin{eqnarray}\label{projaverovertracesallG}
Q^\lambda_{NM}\big(\text{Tr}_{U(N)}(\sigma\Psi)\big)&=&\frac{|S_\lambda|}{|S_n|}\sum_{\alpha\in S_n}\sum_{\substack{R\vdash n\\
\mu\text{ irrep of } S_\lambda \\
i=1,\dots, g(R;\mu)}}\frac{f_R^{U(N)}}{f_\mu^{U(M)}}b_{R,\mu,ii}(\alpha^{-1}\sigma)\text{Tr}_{U(N)}(\alpha\Psi)\nonumber \\
Q^\lambda_{NM}\big(\text{Tr}_{SO(N)}(\sigma\Psi)\big)&=&\frac{|S_\lambda|}{|S_\lambda[S_2]|^2|S_n|}\sum_{\alpha\in S_{2n}}\sum_{\substack{R\vdash n\\
\mu\text{ irrep of } S_\lambda \\
i=1,\dots, g(R;2\mu)}} \frac{f_R^{SO(N)}}{f_\mu^{SO(M)}}b^+_{R,\mu,ii}(\alpha^{-1}\sigma)\text{Tr}_{SO(N)}(\alpha\Psi)\nonumber \\Q^\lambda_{NM}\big(\text{Tr}_{Sp(N)}(\sigma\Psi)\big)&=&\frac{|S_\lambda|}{|S_\lambda[S_2]|^2|S_n|}\sum_{\alpha\in S_{2n}}\sum_{\substack{R\vdash n\\
\mu\text{ irrep of } S_\lambda \\
i=1,\dots, g(R;\mu\cup\mu)}} \frac{f_R^{Sp(N)}}{f_\mu^{Sp(M)}}b^-_{R,\mu,ii}(\alpha^{-1}\sigma)\text{Tr}_{Sp(N)}(\alpha\Psi).\nonumber \\
\end{eqnarray}
As said before, to prove that $Q^\lambda_{NM}$ is self-adjoint, the necessary and sufficient condition is that the result of 
\begin{equation}\label{correlatorwithQ}
\langle (Q^\lambda_{NM}\text{Tr}_{G(N)}(\sigma \Psi)\text{Tr}_{G(N)}(\tau \bar{\Psi})\rangle
\end{equation}
is invariant under the swap $\sigma\leftrightarrow \tau$. This is the case for all gauge groups we are considering. One has to use (\ref{projaverovertracesallG}) in (\ref{correlatorwithQ}), apply the Wick contractions for multitrace monomials 
(\ref{wicksmultitracesunitary}), (\ref{wicksmultitraceorthogonal}) and (\ref{wicksmultitracesymplectic}), and take into account that $\sum_ib^{(\pm)}_{R,\mu,ii}(\sigma)=\sum_ib^{(\pm)}_{R,\mu,ii}(\sigma^{-1})$.\\

\subsection{Eigenvectors and eigenvalues}

For the eigenvectors and eigenvalues of $Q_{NM}^\lambda=\text{Proj}_{NM}\circ \text{Av}_{MN}^{\lambda}$, take the restricted Schur polynomials with our normalization:
\begin{equation*}
\chi^{G(N)}_{R,\mu,m}(\Psi)=\sum_{\sigma\in S_n(S_{2n})}b^{G}_{R,\mu,m}(\sigma)\text{Tr}(\sigma\Psi).
\end{equation*}
Again,  $\sum_ib^{(\pm)}_{R,\mu,ii}(\sigma)=\sum_ib^{(\pm)}_{R,\mu,ii}(\sigma^{-1})$ and the convolution products (\ref{third}) and (\ref{thirdSp}) we see that
\begin{eqnarray}\label{Qlambdageneral}
Q_{NM}^\lambda(\chi^{U(N)}_{R,\mu,ij}(\Psi))&=&\frac{|S_\lambda|}{|S_n|}\frac{f_R^{U(N)}}{f_\mu^{U(M)}}\chi^{U(N)}_{R,\mu,ij}(\Psi) \nonumber \\
Q_{NM}^\lambda(\chi^{SO(N)}_{R,\mu,i}(\Psi))&=&\frac{|S_\lambda|}{|S_\lambda[S_2]|^2|S_{2n}|}\frac{f_R^{SO(N)}}{f_\mu^{SO(M)}}\chi^{SO(N)}_{R,\mu,i}(\Psi) \nonumber \\
Q_{NM}^\lambda(\chi^{Sp(N)}_{R,\mu,i}(\Psi))&=&\frac{|S_\lambda|}{|S_\lambda[S_2]|^2|S_{2n}|}\frac{f_R^{Sp(N)}}{f_\mu^{Sp(M)}}\chi^{Sp(N)}_{R,\mu,i}(\Psi),
\end{eqnarray}
and restricted Schur polynomials are eigenvalues of $Q^\lambda$.\\
Note that if we apply $Q_{NM}^\lambda$ to operators with a different distribution of fields $\lambda'$, say
\begin{equation*}
n'_1+n'_2+\cdots+n'_r=n
\end{equation*}
then $\mu\neq\mu'$ for $\mu$ irrep of $S_\lambda$ and $\mu'$ irrep of $S_{\lambda'}$, and it is 0. This fact allows us to define
\begin{equation*}
Q^{\vdash\vdash n}_{NM}=\sum_{\substack{\lambda\vdash n\\
\lambda\neq (n)}}Q^\lambda_{NM}
\end{equation*}
as charges with the same properties as (\ref{Qlambdageneral}) which act non-trivially on all gauge invariant operators (except for half-BPS) built on $n$ fields. 

\section{Conclusions and future works}
In this paper we have constructed two infinite sets of self-adjoint commuting charges for a quite general CFT. They come out naturally by considering an infinite embedding chain of Lie algebras, an underlying structure that share all theories with gauge groups $U(N)$, $SO(N)$ and $Sp(N)$.  The generality of the construction allows us to carry all gauge groups at the same time in a unified framework, and so to understand the similarities among them.\\
One of the surprising results is that among all the bases of operators which diagonalize the free-field two-point function, restricted Schur polynomials are singled out. They are the eigenstates of the charges. Moreover, the charges, via their eigenvalues, resolve the labels of the restricted Schur polynomials. The correlator of two restricted Schur polynomials can be read (up to constants) from the eigenvalues of the charges as well.\\
We also have suggested that the charges should correspond to asymptotic multipole moments of the geometries in the gravity side although, for obvious reasons we explain in the paper, we were not able to establish an explicit connection. For unitary groups, we have shown that the eigenvalues of the charges admit a probabilistic interpretation in the space of paths of the branching graph of the unitary group.\\
There are a number of future works that this paper suggests. Let us list some of them.
\begin{itemize}
\item Construction of charges $Q^m_{NM}$. This charges will break the degeneracy we still have in the multiplicity labels. The sets $\{Q^{\vdash n}_{NM}\}$, $\{Q^{\vdash \vdash n}_{NM}\}$  and $\{Q^m_{NM}\}$  will complete the specification of state.
\item It is likely that the  eigenvalues of $Q^{\vdash \vdash n}_{NM}$ have a similar probabilistic interpretation as the eigenvalues of  $Q^{\vdash n}_{NM}$. They will be related to some Markov process in the space of paths of branching graphs. It would be interesting to identify it. We also think that these processes must have a physical meaning in the gravity side. It should be investigated.
\item We used matter in the adjoint for the construction of the charges. tt would be interesting to see if we can relax this condition and apply the machinery to theories with matter, say, in the bifundamental.
\item In the same line as the previous point, It would be worth investigating the applicability of our charges to quiver gauge theories and see, for example, if generalized restricted Schurs, as described in \cite{DKN} are their eigenstates.
\item Weingarten calculus is a powerful tool and can be further exploited. For example, it seems possible to rewrite the dilatation operator in terms of Weingarten integrals. Then, perhaps, we can use the properties of those integrals, which are being actively studied, to say something about nonplanar integrability.   

\end{itemize}

\section*{Acknowledgements}
The author would like to thank Robert de Mello Koch whose ideas and suggestions have helped to enrich this paper. The author is also grateful to Vishnu Jejjala, Suresh Nampuri and Alvaro Veliz Osorio for their useful comments. This work has been partly supported by a Claude Leon Fellowship.

\appendix
\section{Algebra with indeces}\label{awi}
Delta tensors
\begin{equation*}
\delta^I_J=\delta^{i_1}_{j_1}\delta^{i_2}_{j_2}\cdots \delta^{i_n}_{j_n}
\end{equation*}
are easily seen to fulfill
\begin{equation}\label{basicprop}
\delta^{\alpha(I)}_{\alpha(J)}=\delta^{i_{\alpha(1)}}_{j_{\alpha(1)}}\cdots \delta^{i_{\alpha(1)}}_{j_{\alpha(1)}}=\delta^I_J.
\end{equation}
This happens because deltas commute with each other. Applying (\ref{basicprop}) we see that
\begin{equation*}
(\alpha)^I_J\equiv \delta^I_{\alpha(J)}=\delta^{\alpha^{-1}(I)}_J=(\alpha^{-1})^J_I.
\end{equation*}
When we have different permutations up and downstairs one can see that
\begin{equation*}
\delta^{\alpha(I)}_{\beta(J)}=\delta^I_{\beta\alpha^{-1}(J)}=(\beta\alpha^{-1})^I_K.
\end{equation*}
We often have to perform products like
\begin{equation*}
(\alpha)^I_J(\beta)^J_K=\delta^I_{\alpha(J)}\delta^J_{\beta(K)}=\delta^I_J\delta^{\alpha^{-1}(J)}_{\beta(K)}=\delta^I_J\delta^J_{\beta\alpha(K)}=(\beta\alpha)^I_K.
\end{equation*}
For $SO(N)$ gauge group we often need 
\begin{equation*}
\delta_{\alpha(I)}\delta^{\beta(I)}=\delta_K\delta_{K'}(\alpha)^K_I(\beta)^{K'}_I=\delta_K\delta_{K'}(\alpha^{-1})^I_K(\beta)^{K'}_I=\delta_I\delta^{\alpha^{-1}\beta(I)}=\text{Tr}_{SO(N)}(\alpha^{-1}\beta).
\end{equation*}
Different rules apply when we deal with tensors that have other kind of symmetry like $\Psi$ and $\big[g\big]$.
Tensors $\Psi$ (and $\big[g\big]$)  have the obvious symmetry
\begin{equation*}
\Psi^{\alpha(I)}_{\alpha(J)}=\Psi^I_J, \quad \alpha \in S_\lambda,
\end{equation*}
but it is not true for $\alpha\notin S_n\times S_m$. A direct consequence of this fact is that, for generic $\alpha\in S_{n}$, 
\begin{equation*}
\Psi^{\alpha(I)}_J\neq \Psi^I_{\alpha^{-1}(J)}.
\end{equation*}

\section{Shuffling slots for $\lambda$-adjoint actions}\label{GAA}
Generic multitrace monomials for unitary gauge theories of a total number of $n$ fields can be written in terms of $\sigma\in S_n$. In doing so, we have tacitly chosen a given order of fields. Let us choose some $\Psi$ with $\lambda$-structure
\begin{equation*}
\text{Tr}(\sigma\Psi)\equiv \Psi^J_{J'} (\sigma)^{J'}_J,\quad \sigma\in S_n.
\end{equation*} 
Imagine we want to write the same multitrace monomial in terms of the tensor $\Psi^{\alpha(J)}_{J'}$, for $\alpha\in S_n$. It is clear that there should be a $\sigma'$ different from $\sigma$ that encodes the same monomial, that is,
\begin{equation*}
\Psi^{\alpha(J)}_{J'}(\sigma')^{J'}_J=\Psi^J_{J'} (\sigma)^{J'}_J.
\end{equation*}
But
\begin{equation*}
\Psi^{\alpha(J)}_{J'}(\sigma')^{J'}_J=\Psi^{J}_{J'}(\sigma')^{J'}_{\alpha^{-1}(J)}=\Psi^{J}_{J'}(\alpha^{-1}\sigma')^{J'}_J,
\end{equation*}
so, $\sigma'=\alpha\sigma$ referred to $\Psi^{\alpha(J)}_{J'}$ drives the same multitrace monomial as $\sigma$ does when referred to $\Psi^J_{J'}$. We have
\begin{equation}\label{shuffling}
\Psi^{\alpha(J)}_{J'}(\alpha\sigma)^{J'}_J=\Psi^J_{J'} (\sigma)^{J'}_J.
\end{equation}
Similarly,
\begin{equation*}
\Psi^{J}_{\beta(J')}(\sigma\beta^{-1})^{J'}_J=\Psi^J_{J'} (\sigma)^{J'}_J.
\end{equation*}
Now, when applying the naive adjoint action we transform
\begin{equation*}
\Psi^J_{J'} (\sigma)^{J'}_J \to \big[g\big]^I_J \Psi^J_{J'} \big[\bar{g}\big]^{I'}_{J'}(\sigma)^{I'}_I.
\end{equation*}
But the correct action needs to shuffle the group elements $g_1,\dots,g_r \in U(N)$. So we are interested in
\begin{equation*}
\big[g\big]^I_{\alpha(J)} \Psi^J_{J'} \big[\bar{g}\big]^{I'}_{J'}(\sigma)^{I'}_I=\big[g\big]^I_J \Psi^{\alpha^{-1}(J)}_{J'} \big[\bar{g}\big]^{I'}_{J'}(\sigma)^{I'}_I.
\end{equation*}
We see that $\sigma$ drives a multitrace monomial but refered to the tensor $\Psi^{\alpha^{-1}(J)}_{J'}$, what means that in any shuffling we are changing the multitrace structure. To remedy this we use (\ref{shuffling}) and write
\begin{equation*} 
\big[g\big]^I_{\alpha(J)} \Psi^J_{J'} \big[\bar{g}\big]^{I'}_{J'}(\alpha^{-1}\sigma)^{I'}_I
\end{equation*}
for every shuffling. Then the generalized adjoint action is defined as
\begin{equation*}
\text{Ad}^{\lambda}_{g}[\text{Tr}_{U(N)}(\sigma \Psi]=\frac{1}{|S_n|}\sum_{\alpha \in S_n} \big[g\big]^I_{\alpha(J)} \Psi^J_{J'} \big[\bar{g}\big]^{I'}_{J'}(\alpha^{-1}\sigma)^{I'}_I.
\end{equation*}

\end{document}